\documentclass[twocolumn,superscriptaddress,floatfix,aps,prd,nofootinbib,preprintnumbers]{revtex4}
\usepackage[colorlinks = true,
            linkcolor = blue,
            urlcolor  = blue,
            citecolor = blue,
            anchorcolor = blue]{hyperref}

\usepackage{aas_macros}
\usepackage{amsmath,amssymb,amsfonts}
\usepackage[dvipdfmx]{graphicx}
\usepackage{color}
\usepackage[T1]{fontenc} 
\usepackage{ulem}

\newcommand{\bfk}{{\boldsymbol k}}
\newcommand{\bfp}{{\boldsymbol p}}

\newcommand{\bfr}{{\boldsymbol r}}
\newcommand{\bfs}{{\boldsymbol s}}
\newcommand{\bfv}{{\boldsymbol v}}
\newcommand{\bfx}{{\boldsymbol x}}

\newcommand{\deltaK}{\delta^{\rm K}}
\newcommand{\deltag}{\delta_{\rm g}}

\newcommand{\deltagz}{\delta_{\rm g}^{\rm(S)}}

\newcommand{\gammaz}{\gamma^{\rm (S)}}

\newcommand{\hatk}{\hat{k}}
\newcommand{\hatp}{\hat{p}}

\newcommand{\bK}{b_{\rm K}}

\newcommand{\pkred}{P^{\rm (S)}}

\begin{document}
\title{Improving redshift-space power spectra of halo  intrinsic alignments from perturbation theory}

\author{Atsushi Taruya}
\affiliation{Center for Gravitational Physics and Quantum Information, Yukawa Institute for Theoretical Physics, Kyoto University, Kyoto 606-8502, Japan}
\affiliation{Kavli Institute for the Physics and Mathematics of the Universe (WPI), The University of Tokyo Institutes for Advanced Study, The University of Tokyo, 5-1-5 Kashiwanoha, Kashiwa, Chiba 277-8583, Japan}
\author{Toshiki Kurita}
\affiliation{Max-Planck-Institut f$\ddot{u}$r Astrophysik, Karl-Schwarzschild-Strasse 1, 85741 Garching, Germany}
\affiliation{Kavli Institute for the Physics and Mathematics of the Universe (WPI), The University of Tokyo Institutes for Advanced Study, The University of Tokyo, 5-1-5 Kashiwanoha, Kashiwa, Chiba 277-8583, Japan}
\author{Teppei Okumura}
\affiliation{Academia Sinica Institute of Astronomy and Astrophysics (ASIAA), No. 1, Section 4, Roosevelt Road, Taipei 106216, Taiwan}
\affiliation{Kavli Institute for the Physics and Mathematics of the Universe (WPI), The University of Tokyo Institutes for Advanced Study, The University of Tokyo, 5-1-5 Kashiwanoha, Kashiwa, Chiba 277-8583, Japan}

\date{\today}
\begin{abstract}
Intrinsic alignments (IAs) of galaxies/halos observed via galaxy imaging survey, combined with redshift information, offer a novel probe of cosmology as a tracer of the tidal force field of a large-scale structure. In this paper, 
we present a perturbation theory based model for the redshift-space power spectra of galaxy/halo IAs that can keep the impact of the Fingers-of-God damping effect, known as a nonlinear systematics of redshift-space distortions, under control. Focusing particularly on galaxy/halo density and an IA cross power spectrum, we derive analytically the explicit expressions for the next-to-leading order corrections. Comparing the model predictions with $N$-body simulations, we show that these corrections indeed play an important role for an unbiased determination of the growth-rate parameter, and hence the model proposed here can be used for a precision test of gravity on cosmological scales.  
\end{abstract}
\preprint{YITP-24-110}

\maketitle
\flushbottom
\section{Introduction}
\label{sec:intro}

The large-scale structure of the Universe, as partly traced by galaxy redshift surveys, offers wealth of cosmological information from which one can scrutinize not only the cosmic expansion history but also the growth of structure over the last billion years. Since the establishment of the concordance  cosmological model ($\Lambda$ cold dark matter model, $\Lambda$CDM) in the early 2000s (\cite{BOOMERanG2000,Percival_etal2001,Spergel:2003cb}, see also Refs.~\cite{Riess_etal1998,Perlmutter_etal1999} for distant supernova observations in late 1990s), observations have made significant progress, and the ``stage IV'' dark energy experiments, as defined in the report from the Dark Energy Task Force \cite{DETF_2006}, have at last started (e.g., \cite{DESI_1yr_BAO2024,Euclid_mission2024}). Their dataset will dramatically improve our understanding of the Universe, and definitely help in clarifying the unresolved mysteries on the nature of dark energy and dark matter. In particular, the large-scale structure observations would provide an important clue to the tensions raised by late-time observations against the cosmic microwave background experiments (e.g., \cite{Planck2018_VI_2020,Riess_etal2018,Riess_etal2019,Freedman_etal2019,Freedman_etal2020} for $H_0$ tension, \cite{Hikage_etal2019,DES_1yr2018,KiDS-1000_2021} for $S_8$ tension; see Ref.~\cite{Perivolaropoulos_Skara2022} for a review), and in this respect, an efficient statistical analysis method as well as a sophisticated theoretical modeling need to be developed toward a more stringent constraint on cosmological parameters. 

In maximizing the cosmological information from the galaxy survey data, one promising way would be to utilize not only the positional information of each galaxy  (i.e., angular position on the celestial sphere and redshift) but also to combine the shape or orientation of galaxy images (\cite{Troxel_Ishak2015,Joachimi_etal2015,Kirk_etal2015,Kiessling_etal2015,Lamman_etal2024} for reviews). Indeed, a conventional way to study cosmology with the galaxy redshift surveys has been made mostly with the galaxy positional information alone (i.e., angular position on the celestial sphere and redshift) to evaluate the density fluctuations. On the other hand, the galaxy shape information has been utilized to measure the weak gravitational lensing effect, but their intrinsic contribution, often referred to as the intrinsic alignment (IA), is considered as the systematics to be removed from the cosmological analysis (e.g., \cite{Heavens_etal2000,Lee_Pen2000,Croft_Metzler2000}). Nevertheless, as it has been studied in Refs.~\cite{Schmidt_Chisari_Dvorkin2015,Taruya_Okumura2020,Okumura_Taruya2021,Chuang_Okumura_Shirasaki2021} (see  \cite{Croft_Metzler2000,Okumura_Jing_Li2009,Okumura_Jing2009,Faltenbacher_etal2012} for earlier works), the IA of galaxies in three-dimensional space carries ample cosmological information, and can be used as a tracer of the gravitational tidal field of the large-scale structure. The redshift-space distortions and the baryon acoustic oscillations, as primary targets of the stage-IV galaxy surveys, have been successfully measured at high statistical significance from the statistics of IA \cite{Okumura_Taruya2023,Xu_etal2023}. Combining them with the conventional galaxy clustering data, a tighter constraint on the growth rate of structure was also obtained. Further, a measurement of IA in three dimensions is expected to be beneficial in detecting imprints of the primordial Universe that is even difficult with galaxy clustering data. Examples include gravitational wave background \cite{Schmidt_Jeong2012a,Schmidt_Pajer_Zaldarriaga2014,Biagett_Orlando2020,Akitsu_Li_Okumura2023,Okumura_Sasaki2024}, anisotropic primordial non-Gaussianity \cite{Schmidt_Chisari_Dvorkin2015,Chisari_etal2016,Kogai_etal2018,Kogai_etal2021,Akitsu_etal2021,Kurita_Takada2023}, statistical anisotropies \cite{Shiraishi_Okumura_Akitsu2023}, and primordial magnetic fields \cite{Philcox_etal2024,Saga_Shiraishi_Akitsu_Okumura2024}.

Motivated by these recent works on the IA as a novel cosmological probe, an improved theoretical modeling of IA statistics is an important and crucial issue to put forward further the practical application to observations. This is exactly the aim of this paper. In particular, a perturbation theory modeling of the IA statistics in redshift space is our  main focus. In Ref.~\cite{Okumura_Taruya_Kurita_Nishimichi2024}, extending the work by Ref.~\cite{Okumura_Taruya2019,Okumura_Taruya_Nishimichi2020}, in which the analytical formulas for anisotropic IA correlations are derived based on the linear alignment model \cite{Catelan_etal2001,Hirata_Seljak2004}, we have presented an analytical model of nonlinear correlators of galaxy/halo ellipticities. These predictions are applied to the ellipticity correlation functions measured from the galaxy samples of the Sloan Digital Sky Survey (SDSS) and SDSS-III BOSS surveys, and the first constraints on the growth rate of the Universe, $f(z)\sigma_8(z)$ were derived over the redshift range of $0.16<z<0.7$ \cite{Okumura_Taruya2023}. Combining them with the conventional galaxy clustering statistics (i.e., the autocorrelation function of galaxy density fields), the constraints were shown to get tighter, achieving an improvement of up to $19\%$.

With the improved precision from stage IV surveys, the statistical errors on cosmological parameters will be reduced, and unbiased parameter estimation will become one of the greatest concerns. Similar to the conventional clustering statistics, nonlinear systematics in the IA statistics arise mainly from the gravitational evolution, redshift-space distortions and shape bias, all of which need to be controlled. This indeed explains why, recently, numerous works have appeared to improve the theoretical description of IA based on the perturbation theory treatment (see Ref.~\cite{Bernardeau:2001qr} for a review) and numerical simulations (e.g., Refs.~\cite{Blazek_etal2019,Vlah_Chisari_Schmidt2019,Vlah_Chisari_Schmidt2021,Bakx_etal2023,Matsubara2022a,Matsubara2022b,Matsubara2023,Matsubara2024,Chen_Kokron2024,Maion_etal2024}). Along the line of this, we shall develop a perturbation theory description of the IA statistics in redshift space, taking into account all nonlinear effects mentioned above. As a first step, in this paper, we present an improved prescription of redshift-space cross power spectrum between galaxy/halo density and ellipticity fields, which we shall refer to as GI correlation or GI cross spectrum. As it has been shown in Ref.~\cite{Okumura_Taruya_Kurita_Nishimichi2024}, 
the nonlinearity of the redshift-space distortions, known as the Fingers-of-God (FoG) effect, impacts the IA statistics, leading to the scale-dependent modifications to the power spectra. In particular, the effect is known to give a significant impact on the statistics involving the density fields, and this is why we consider first the GI power spectrum in our series of works. 

Here, following the previous works developed in the conventional galaxy clustering statistics \cite{Taruya:2010mx,Nishimichi:2011jm}, we shall present a general perturbation theory (PT) framework to better control the FoG effect, and compute explicitly the one-loop level (next-to-leading order) predictions of the GI spectra, providing also the analytical formulas for correction terms which are to be added to the standard PT predictions (e.g., Refs.~\cite{Makino:1991rp,Scoccimarro:1996se,Jeong:2006xd}). We then test our predictions against the halo IA statistics in cosmological $N$-body simulations, demonstrating unbiased parameter estimation.  This paper is organized as follows. In Sec.~\ref{sec:tracer_fields}, we begin by describing the redshift-space observables as a tracer of large-scale structure and their relations to the real-space quantities. Section~\ref{sec:power_spectra} introduces the statistics of IA, and starting with their exact expressions, we present a PT model of their statistics, which is derived based on a perturbative expansion of the velocity correlation while keeping the contribution responsible for the FoG effect nonperturbative. In Sec.~\ref{sec:analytical_expressions}, adopting the linear bias relations in density and ellipticity fields, we further elaborate the PT calculations to derive the analytical formulas of the GI cross power spectrum relevant for the next-to-leading order calculations. The explicit expressions of the higher-order correction, the $A_{ij}$ and $B_{ij}$ terms and their projected counterpart, are summarized in Appendixes~\ref{appendix:formulas_Aterm_Bterm} and \ref{appendix:formulas_AEterm_BEterm}. 
Section~\ref{sec:testing_model} discusses the validity of the PT modeling, focusing especially on the GI cross spectrum. 
We perform the parameter inference using the halo catalogs in cosmological $N$-body simulations. In doing so, we adopt the analytical covariance matrix derived from linear theory, whose validity is discussed in Appendix~\ref{appendix:validity_covariance}. Finally, Sec~\ref{sec:conclusion} is devoted to the conclusion and outlook of the present study.

\section{Density and ellipticity fields in redshift space}
\label{sec:tracer_fields}

In this paper, our main focus is the statistics of tracer fields of large-scale matter fluctuations. In particular, we are interested in the galaxy/halo density and the intrinsic shape fields as cosmological 
probes, which we respectively denote by $\delta_{\rm g}$ and $\gamma_{ij}$. The former is measured from the local number density of galaxies/halos $n_{\rm g}$, and the latter is defined by the traceless part of the inertia tensor of galaxy/halo shape, $I_{ij}$\footnote{There are certainly various ways to define the inertia tensor $I_{ij}$. One simple definition would be to express it as $I_{ij}\propto \sum_p w_p\,\Delta x_{p,i}\Delta x_{p,j}$ at locations $\bfx_p$, where $\Delta \bfx_p$ is the position vector of each point centered at $\bfx_p$ (e.g., Refs.~\cite{Kurita_etal2021,Vlah_Chisari_Schmidt2019}). The function $w_p$ represents some weight function given at each position. Various choices of weight functions are used in the actual measurements and simulations, and this would alter the statistical properties of $\gamma_{ij}$. Nevertheless, in perturbative approach employing the parameterized prescription to relate between ellipticity and matter density fields, the impact of weight functions can be, in general, absorbed into the coefficients of perturbative corrections.}. In real space, they are given as 
functions of Eulerian position $\bfx$ as  
\begin{align}
\delta_{\rm g}(\bfx)&=\frac{n_{\rm g}(\bfx)-\overline{n}_{\rm g}}{\overline{n}_{\rm g}},
\label{eq:def_of_deltag}
\\
\gamma_{ij}(\bfx)&=\frac{I_{ij}(\bfx)-\frac{1}{3}\deltaK_{ij}\,{\rm Tr}\,I_{k\ell}(\bfx)}{\overline{{\rm Tr}\,\,I_{k\ell}}},
\label{eq:def_of_gamma_ij}
\end{align}
where the barred quantities imply those averaged over the survey at a given redshift. The operator ${\rm Tr}$ stands for taking the trace part of a matrix, i.e., ${\rm Tr} I_{k\ell}\equiv \sum_{k}I_{kk}$. In the above, the field $\gamma_{ij}$ characterizes three-dimensional shape, and subscripts $i,\,j$ run over $x$, $y$ and $z$. Although the actual intrinsic shape or IA of galaxies that one can observe is a two-dimensional projection onto the sky, we shall consider it as the three-dimensional tensor field for the sake of generality, and later evaluate its projection.

In galaxy redshift surveys, the tracer fields in Eqs.~(\ref{eq:def_of_deltag}) and \eqref{eq:def_of_gamma_ij} are in fact the observables in redshift space, and are given as a function of redshift-space position $\bfs$. Throughout the paper, we work with the distant-observer or plane-parallel limit, where the line-of-sight direction is fixed to the $z$ axis as a specific direction. Then, the redshift-space position $\bfs$ is mapped from the real-space position $\bfx$ through
\begin{align}
 \bfs=\bfx + \frac{1}{a\,H(a)}\,v_z(\bfx)\,\hat{z}
\label{eq:red_real_space}
\end{align}
with the quantities $a$ and $H$ being respectively the scale factor of the Universe and the Hubble parameter at redshift $z\equiv 1/a-1$. 
The quantity $\hat{z}$ is defined as the unit vector parallel to the $z$ axis, and the field $v_z$ is the line-of-sight component of the peculiar velocity field $\bfv$, defined by $v_z=\bfv\cdot\hat{z}$. 

With this relation at Eq.~\eqref{eq:red_real_space}, the real-space galaxy/halo density field $\delta_{\rm g}$ is mapped into redshift-space density field $\delta_{\rm g}^{\rm(S)}$ as follows (e.g., \cite{Scoccimarro:2004tg, Taruya:2010mx}): 
\begin{align}
1+ \deltagz(\bfs) &= \Bigl|\frac{\partial \bfs}{\partial \bfx}\Bigr|^{-1}\bigl\{1+\deltag(\bfx)\bigr\}\Biggr|_{\bfs}
\nonumber
\\
&= \int\frac{d^3\bfk}{(2\pi)^3}\,\int d^3\bfx\,e^{i\,\bfk\cdot(\bfs-\bfx+f\,u_z\hat{z})}\bigl\{1+\deltag(\bfx)\bigr\}, 
\label{eq:deltag_in_config}
\end{align}
which can be recast in Fourier space as\footnote{An intermediate step from Eq.~\eqref{eq:deltag_in_config} to Eq.~\eqref{eq:deltag_in_Fourier} is given below:
\begin{align}
 \deltagz(\bfk)
 &=\int d^3\bfs \,e^{-i\,\bfk\cdot\bfs}
 \Biggl[\Bigl|\frac{\partial\bfs}{\partial\bfx}\Bigr|^{-1}\bigl\{1+\deltag(\bfx)\bigr\}-1\Biggr]
 \nonumber
 \\
 &=\int d^3\bfx \,e^{-i\,\bfk\cdot\{\bfx-f\,u_z\,\hat{z}\}}\Bigl\{1 + \deltag(\bfx)\Bigr\} -\int d^3\bfs \,e^{-i\,\bfk\cdot\bfs}
\nonumber
\\
&=\int d^3\bfx \,e^{-i\,\bfk\cdot\{\bfx-f\,u_z\,\hat{z}\}} 
\Bigl\{1 + \deltag(\bfx) -
\Bigl|\frac{\partial\bfs}{\partial\bfx}\Bigr|
\Bigr\}.
\nonumber
\end{align}
Then, plugging the Jacobian $|\partial\bfs/\partial\bfx|=1-f\nabla_zu_z$ into the above equation yields Eq.~\eqref{eq:deltag_in_Fourier}. 
}
\begin{align}
 \deltagz(\bfk)=\int d^3\bfx \,e^{-i\,\bfk\cdot\{\bfx-f\,u_z\,\hat{z}\}}\Bigl\{\deltag(\bfx)+f\,\nabla_zu_z(\bfx)\Bigr\},
\label{eq:deltag_in_Fourier}
\end{align}
where we define the normalized line-of-sight (infall) velocity  $u_z\equiv-(\bfv\cdot\hat{z})/(a\,H\,f)$. The factor $f$ represents the 
logarithmic 
linear growth rate defined by $f\equiv \ln D_+(z)/d\ln a$ as a key parameter of the test of gravity on cosmological scales. 

Similarly, the galaxy/halo ellipticity field, $\gamma_{ij}$, is mapped into the redshift-space counterpart as\footnote{Here we assume that the tensorial property of $\gamma_{ij}$ is determined by the internal properties of galaxy/halo, and is not affected by the coordinate transformation at Eq.~\eqref{eq:red_real_space}. } 
\begin{align}
 \gammaz_{ij}(\bfk)=\int d^3\bfx \,e^{-i\,\bfk\cdot\{\bfx-f\,u_z\,\hat{z}\}}\,\gamma_{ij}(\bfx).
\label{eq:gamma_in_Fourier}
\end{align}
Notice that the ellipticity field $\gamma_{ij}$ is the density-weighted traceless tensor field. Both the density and ellipticity fields are tracer fields, and with a general prescription of the galaxy density/shape bias \cite{Desjacques_Jeong_Schmidt2018,Vlah_Chisari_Schmidt2019}, they are related perturbatively to the mass density fields $\delta$ in terms of local bias operators.

\section{Modeling redshift-space power spectra}
\label{sec:power_spectra}

In this section, given the redshift-space tracer fields, $\delta^{\rm (S)}_{\rm g}$ and $\gamma_{ij}^{\rm(S)}$, we consider the PT based model of the redshift-space power spectra.

\subsection{Power spectra of galaxy/halo intrinsic alignments}
\label{subsec:IA_power_spectra}

Combining the galaxy/halo density and ellipticity fields as our basic  observables, we measure three power spectra, i.e., autopower spectra of galaxy/halo density and ellipticity fields and their cross power spectrum. Among them, the spectra related to the IA are defined by 
\begin{align}
& \langle \deltagz(\bfk)\gammaz_{ij}(\bfk')\rangle=(2\pi)^3\delta_{\rm D}(\bfk+\bfk')\,\pkred_{{\rm g\gamma},ij}(\bfk),
\label{eq:pkred_g_ij}
\\
& \langle\gammaz_{ij}(\bfk)\gammaz_{kl}(\bfk')\rangle=(2\pi)^3\delta_{\rm D}(\bfk+\bfk')\,\pkred_{\gamma\gamma,ijkl}(\bfk),
\label{eq:pkred_ijkl}
\end{align}
with the bracket $\langle\cdots\rangle$ indicating the ensemble average over the randomness of initial conditions. We shall denote the former by the GI power spectrum. The latter is often called the II power spectrum. 

Plugging the expressions at Eqs.~\eqref{eq:deltag_in_Fourier} and \eqref{eq:gamma_in_Fourier} into the above, the redshift-space power spectra are rewritten as follows:
\begin{align}
 \pkred_{{\rm g},ij}(\bfk)&=\int d^3\bfr \,e^{-i\,\bfk\cdot\bfr}\,\Bigl\langle e^{i\,k\mu\,f\Delta u_z}\,
\nonumber
\\
&\quad\times
 \Bigl\{\deltag(\bfx)+f\,\nabla_zu_z(\bfx)\Bigr\}\,
\gamma_{ij}(\bfx')\Bigr\rangle,
\label{eq:pkred_gal_ij}
\\
 \pkred_{\gamma\gamma,ijkl}(\bfk)&=\int d^3\bfr \,e^{-i\,\bfk\cdot\bfr}\,
\Bigl\langle e^{i\,k\mu\,f\Delta u_z}\, \gamma_{ij}(\bfx)\gamma_{kl}(\bfx')
\Bigr\rangle,
\label{eq:pkred_gam_gam_ijkl}
\end{align}
where we define $\bfr\equiv \bfx-\bfx'$ and $\Delta u_z\equiv u_z(\bfx)-u_z(\bfx')$.  The quantity $\mu$ is the directional cosine defined by $\mu=\bfk\cdot\hat{z}/|\bfk|$. 
In deriving these expressions, we assume that the two-point statistics enclosed by brackets above depend only on the separation vector, $\bfx-\bfx'$. In Eqs.~\eqref{eq:pkred_gal_ij} and \eqref{eq:pkred_gam_gam_ijkl}, we do not consider the fields  $\deltag$, $u_z$, and $\gamma_{ij}$ to be small and perturbative quantities. In this sense, they are the nonperturbative expressions for the IA power spectra. Starting with these expressions, we derive an analytically tractable but properly accurate description of these spectra, which can be dealt with perturbation theory treatment.

Although our main focus in this paper is to develop a PT model of the GI spectrum $P^{\rm (S)}_{\rm g\gamma,ij}$ and to present explicit expressions for the next-to-leading order PT results, the derivation of the PT model for the power spectrum $P^{\rm (S)}_{\gamma\gamma,ijkl}$ is straightforward and is almost parallel to that for the GI spectrum. Hence, we also discuss the PT model of $P^{\rm (S)}_{\gamma\gamma,ijkl}$, leaving a detailed calculation and comparison with simulations to future work.

\subsection{Perturbation theory based modeling}
\label{subsec:PT_model}

Our starting point is to rewrite Eqs.~\eqref{eq:pkred_gal_ij} and \eqref{eq:pkred_gam_gam_ijkl} in terms of cumulants. Since the structure of the power spectrum expressions given above resemble each other, we shall first consider the cross spectrum $\pkred_{{\rm g\gamma},ij}$ to derive a PT model in Sec.~\ref{subsubsec:GI_cross_spectrum}. We then present the resultant expressions for $\pkred_{\gamma\gamma,ijkl}$ in Sec.~\ref{subsubsec:II_auto_spectrum_1}.  

\subsubsection{GI cross power spectrum, $\pkred_{{\rm g\gamma},ij}$}
\label{subsubsec:GI_cross_spectrum}

Let us define building blocks of the power spectrum below:  
\begin{align}
j_1 &= i k\mu f,
\label{eq:building_block1}
\\
A_1 &= \Delta u_z,
\label{eq:building_block2}
\\
A_2 &= \deltag(\bfx) + f\nabla_zu_z(\bfx),
\label{eq:building_block3}
\\
A_3 &= \gamma_{ij}(\bfx').
\label{eq:building_block4}
\end{align}
Equation~\eqref{eq:pkred_gal_ij} is then rewritten with
\begin{align}
 \pkred_{{\rm g\gamma},ij}(\bfk)&=\int d^3\bfr\, e^{-i\,\bfk\cdot\bfr}\Bigl\langle
e^{j_1A_1}\,A_2 A_3\Bigr\rangle.
\label{eq:pkred_GI}
\end{align}
In the above, the difficulty in analytical calculations arises from the exponent in the brackets, which, when Taylor expanded, generates an infinite series of multipoint spectra involving the density, ellipticity, and velocity fields. In order to find an analytically tractable expression while keeping a part of non-perturbative properties, we rewrite the ensemble average in the integrand in terms of cumulants, following Refs.~\cite{Taruya:2010mx,Taruya:2013my}. The cumulant correlation quantifies the connected part of multipoint correlations, isolating non-Gaussian features in a statistical field and thus simplifying the statistical description of complex fields.

In terms of the cumulants denoted by $\langle\cdots\rangle_{\rm c}$, the bracket in the integrand is expressed as\footnote{The first equality in Eq.~\eqref{eq:cumulant_expansion} comes from the relation that generally holds between cumulant and moment generating functions for a set of fields. Grouping the fields $A_1$, $A_2$, and $A_3$ as ${\boldsymbol A}=\{A_1,A_2,A_3\}$, the relation is given by (e.g., \cite{Scoccimarro:2004tg,Matsubara2008a,Taruya:2010mx,Taruya:2013my})
\begin{align}
\langle e^{{\boldsymbol j}\cdot{\boldsymbol A} }\rangle=\exp\langle e^{{\boldsymbol j}\cdot{\boldsymbol A}}\rangle_{\rm c}
\nonumber
\end{align}
with ${\boldsymbol j}$ being a arbitrary constant vector, ${\boldsymbol j}=\{j_1,j_2,j_3\}$. Taking the derivative twice with respect to $j_2$ and $j_3$, and then setting $j_2=j_3=0$, we obtain the first equality in Eq.~\eqref{eq:cumulant_expansion}.
} 
\begin{align}
 \Bigl\langle
&e^{j_1A_1}\,A_2 A_3\Bigr\rangle 
\nonumber
\\
&\quad = \exp\bigl\langle e^{j_1A_1}\bigr\rangle_{\rm c} 
\nonumber
\\
&\qquad\times\Bigl\{ \bigl\langle e^{j_1A_1}A_2A_3\bigr\rangle_{\rm c} + 
\bigl\langle e^{j_1A_1}A_2\bigr\rangle_{\rm c}\bigl\langle e^{j_1A_1}A_3\bigr\rangle_{\rm c}
\Bigr\}
\nonumber
\\
&\quad\simeq  \exp\bigl\langle e^{j_1A_1}\bigr\rangle_{\rm c} \Bigl[
\bigl\langle A_2A_3\bigr\rangle_{\rm c} 
+ j_1 \,\langle A_1A_2A_3\rangle_{\rm c} 
\nonumber
\\
&\qquad + j_1^2\Bigl\{
\bigl\langle A_1A_2\bigr\rangle_c\bigl\langle A_1A_3\bigr\rangle_{\rm c} 
+\frac{1}{2}\langle A_1^2A_2A_3\rangle_{\rm c}
\Bigr\}+\cdots
\Bigr], 
\label{eq:cumulant_expansion}
\end{align}
where in the second equality, we partly expand the exponential factor in the braces, assuming that $|j_1 A_1|$ is parturbatively small. Here we suppose that the zero mean is ensured for the quantities $A_k$. 

In the expression given above, we do not perturbatively expand  
the exponential prefactor, $\exp\langle  e^{j_1A_1}\rangle$. As 
it has been discussed in Ref.~\cite{Taruya:2010mx} (see also Refs.~\cite{Scoccimarro:2004tg,Taruya:2013my,Hashimoto_etal2017}), in contrast to those expanded in the bracket, this factor involves a zero-lag velocity correlation, and is prone to be dominated by the virialized random motion, leading to the suppression of the power spectrum. This suppression is referred to as the Fingers-of-God effect \cite{Jackson1972,Davis_Peebles1983,Scoccimarro:2004tg}. In the autopower spectrum of galaxy/halo density fields, the suppression appears broadband, and is known to be significant even at BAO scales. Since this would not be properly captured by the perturbative expansion, we may treat it as nonperturbative. Rather, following Refs.~\cite{Taruya:2010mx,Taruya:2013my,Hashimoto_etal2017}, we replace it with a  general functional form, $D_{\rm FoG}(k\mu f\sigma_{\rm v})$, with $\sigma_{\rm v}$ being a constant\footnote{This proposition implies that the spatial correlation in the exponential prefactor is ignored and the zero-lag correlation is only considered. As it has been discussed in Ref.~ \cite{Taruya:2010mx}, the spatial correlation leads to a monotonic change in the broad band power, and does not alter the BAO features drastically. Hence, for the scales of our interest, its impact can be effectively absorbed into the damping function $D_{\rm FoG}$ by varying the parameter $\sigma_{\rm v}$.}. The relevant functional form of $D_{\rm FoG}$ as well as the choice of constant $\sigma_{\rm v}$ will be discussed later. With the proposition mentioned above, plugging Eq.~\eqref{eq:cumulant_expansion} into the power spectrum expression at Eq.~\eqref{eq:pkred_GI} leads to 
\begin{align}
 \pkred_{{\rm g\gamma},ij}(\bfk)&\simeq \int d^3\bfr\, e^{-i\,\bfk\cdot\bfr}
D_{\rm FoG}(k\mu f\sigma_{\rm v}) 
\Bigl[
\bigl\langle A_2A_3\bigr\rangle_c 
\nonumber
\\
& + j_1 \langle A_1A_2A_3\rangle_c + 
j_1^2\bigl\langle A_1A_2\bigr\rangle_c\bigl\langle A_1A_3\bigr\rangle_c \Bigr].
\label{eq:pkred_GI_model}
\end{align}
Here, in the bracket, we only retain the cumulant correlators relevant at one-loop order of PT calculations. The structure of Eq.~\eqref{eq:pkred_GI_model} resembles the expression first derived in Ref.~\cite{Taruya:2010mx} for the autopower spectrum of galaxy/halo density fields.

For more explicit expressions relevant to the PT calculations, we define the real-space power spectra:
\begin{align}
 \langle \deltag(\bfk)\gamma_{ij}(\bfk')\rangle &= (2\pi)^3\delta_{\rm D}(\bfk+\bfk') \,P_{{\rm g\gamma},ij}(\bfk),
 \label{eq:power_deltag_gamma}
\\
 \langle \theta(\bfk)\gamma_{ij}(\bfk')\rangle &= (2\pi)^3\delta_{\rm D}(\bfk+\bfk') \,P_{{\rm \theta \gamma},ij}(\bfk),
 \label{eq:power_theta_gamma}
\end{align}
where the quantity $\theta$ is the velocity-divergence field, given by $\theta=-\nabla\cdot\bfv/(aH\,f)$. Equation~\eqref{eq:pkred_GI_model} is then rewritten as\footnote{Here, we implicitly assume that the terms in the bracket at Eq~\eqref{eq:pkred_GI_model} can be dealt with PT, and the velocity field is described as an irrotational flow, as similarly made in the autopower spectrum of the galaxy/halo density field in Refs.~\cite{Taruya:2010mx,Taruya:2013my}. }
\begin{align}
\pkred_{{\rm g\gamma},ij} (\bfk)&=D_{\rm FoG}(k\mu f\sigma_{\rm v}) 
\Bigl[P_{{\rm g\gamma},ij}(\bfk)+f\,\mu^2 P_{{\rm \theta\gamma},ij}(\bfk) 
\nonumber
\\
&
+ A_{ij}(\bfk) + B_{ij}(\bfk) \Bigr],
\label{eq:TKO_GI_general}
\end{align}
where the functions $A_{ij}$ and $B_{ij}$ are regarded as higher-order terms that can be ignored in the linear theory limit. These are analogous to the $A$ and $B$ terms in the model of Ref.~\cite{Taruya:2010mx} for the galaxy/halo density power spectrum, defined by 
\begin{align}
 A_{ij}(\bfk)&= k\,\mu f\int \frac{d^3\bfp}{(2\pi)^3}\,\frac{p_z}{p^2}\Bigl\{ 
\tilde{B}_{ij}(\bfp,\bfk-\bfp,-\bfk)
\nonumber
\\
&-\tilde{B}_{ij}(\bfp,\bfk,-\bfk-\bfp).
\Bigr\}
\label{eq:A_ij}
\\
 B_{ij}(\bfk)&= (k\,\mu f)^2 \int \frac{d^3\bfp}{(2\pi)^3}\,F(\bfp)G_{ij}(\bfk-\bfp);
\label{eq:B_ij}
\\
&\qquad
F(\bfp)=\frac{p_z}{p^2}\Bigl\{P_{g\theta}(p)+f\,\frac{p_z^2}{p^2} \,P_{\theta\theta}(p)\Bigr\},
\\
&\qquad
G_{ij}(\bfp)=\frac{p_z}{p^2}\,P_{{\rm \theta \gamma},ij}(\bfp)
\end{align}
with the function $\tilde{B}_{ij}$ being the cross bispectra defined by
\begin{align}
& \Bigl\langle 
\theta(\bfk_1)\Bigl\{\deltag(\bfk_2)+f\,\frac{k_{2z}^2}{k_2^2}\theta(\bfk_2)\Bigr\}\gamma_{ij}(\bfk_3)
\Bigr\rangle
\nonumber
\\
&=(2\pi)^3\delta_{\rm D}(\bfk_1+\bfk_2+\bfk_3)\,\tilde{B}_{ij}(\bfk_1,\bfk_2,\bfk_3).
\label{eq:cross_B_ij}
\end{align}
Equation~\eqref{eq:TKO_GI_general} with Eqs.~\eqref{eq:A_ij}-\eqref{eq:cross_B_ij} are the key expressions for the PT model of the GI cross spectrum. Note here that while we retain the terms relevant to the next-to-leading order PT calculations, the expressions given above are still general, and any treatment of PT calculation, including the resummed or renormalized PT treatments (e.g., Refs.~\cite{Crocce:2005xy,Crocce:2007dt,Valageas2007,Taruya:2007xy,Bernardeau:2008fa,Matsubara2008a,Pietroni:2008jx,Crocce:2012fa,Taruya:2012ut}),  or numerical simulations utilizing the emulator technique (e.g., Refs.~\cite{Heitmann:2006hr,Habib:2007ca}, see Ref.~\cite{Moriwaki_Nishimichi_Yoshida2023} for a review) can be applied to model and predict each term. In this paper, employing the standard PT calculations, we explicitly compute Eq.~\eqref{eq:TKO_GI_general}, and its model prediction is compared with cosmological $N$-body simulations.

\subsubsection{II power spectrum: $\pkred_{\gamma\gamma,ijkl}$}
\label{subsubsec:II_auto_spectrum_1}

Here, we briefly touch on the PT model of the redshift-space II power spectrum $\pkred_{\gamma\gamma,ijkl}$, i.e., the autopower spectrum of the ellipticity fields in redshift space. Starting with the expression at Eq.~\eqref{eq:pkred_gam_gam_ijkl}, we follow the same steps as described in Sec.~\ref{subsubsec:GI_cross_spectrum}, but we replace respectively the quantities $A_2$ and $A_3$ given in Eqs.~\eqref{eq:building_block3} and \eqref{eq:building_block4} with $A_2=\gamma_{ij}(\bfx)$ and $A_3=\gamma_{kl}(\bfx')$. Then, we arrive at the expression similar to Eq.~\eqref{eq:TKO_GI_general}:
\begin{align}
 \pkred_{\gamma\gamma,ijkl}(\bfk)&=D_{\rm FoG}(k\mu f\sigma_{\rm v})
\nonumber
\\
& \times \Bigl[P_{\gamma\gamma,ijkl}(\bfk) + A_{ijkl}(\bfk) +
B_{ijkl}(\bfk)
\Bigr],
\label{eq:TKO_II_general}
\end{align}
where the function $P_{\gamma\gamma,ijkl}$ is the real-space II power spectrum defined by
\begin{align}
 \langle \gamma_{ij}(\bfk) \gamma_{kl}(\bfk') \rangle &= (2\pi)^3\delta_{\rm D}(\bfk+\bfk') \,P_{\gamma\gamma,ijkl}(\bfk).
\end{align}
The terms $A_{ijkl}$ and $B_{ijkl}$ are the next-to-leading order corrections to the linear theory prediction. Their expressions are 
\begin{align}
 A_{ijkl}(\bfk)&= k\,\mu f\int \frac{d^3\bfp}{(2\pi)^3}\,\frac{p_z}{p^2}\Bigl\{ 
\tilde{B}_{ijkl}(\bfp,\bfk-\bfp,-\bfk)
\nonumber
\\
&-\tilde{B}_{ijkl}(\bfp,\bfk,-\bfk-\bfp)
\Bigr\},
\label{eq:A_gx}
\\
 B_{ijkl}(\bfk)&= (k\,\mu f)^2 \int \frac{d^3\bfp}{(2\pi)^3}\,F_{ij}(\bfp)F_{kl}(\bfk-\bfp)\,;
\nonumber
\\
&\quad 
F_{ij}(\bfp)=\frac{p_z}{p^2}\,P_{{\rm \theta\gamma},ij}(\bfp)
\end{align}
with the function $\tilde{B}_{ijkl}$ being the real-space cross bispectra defined by
\begin{align}
& \Bigl\langle 
\theta(\bfk_1)\gamma_{ij}(\bfk_2)\gamma_{kl}(\bfk_3)\Bigr\rangle
\nonumber
\\
&=(2\pi)^3\delta_{\rm D}(\bfk_1+\bfk_2+\bfk_3)\,\tilde{B}_{ijkl}(\bfk_1,\bfk_2,\bfk_3).
\end{align}
Note that in the linear theory limit (this also implies the limit of $D_{\rm FoG}\to1$), the II power spectrum does not receive any correction from the RSD. In other words, no explicit dependence on the linear growth rate $f$ appears in the power spectrum expression \cite{Okumura_Taruya2019}. Going beyond linear order, in Eq.~\eqref{eq:TKO_II_general}, the contributions from RSD are evident in the $A_{ijkl}$ and $B_{ijkl}$ terms, which are respectively proportional to $f$ and $f^2$. Although the significance of these terms depends on how the ellipticity field is coupled with the velocity field, an important point to note is that Eq.~\eqref{eq:TKO_II_general} shares a common factor $D_{\rm FoG}$ with the GI cross spectrum, originating from the exponential prefactor, $\exp\langle e^{j_1A_1}\rangle$. This suggests that the II autospectrum is similarly affected by the Fingers-of-God effect (see also Refs.~\cite{Okumura_Taruya2023,Okumura_Taruya_Kurita_Nishimichi2024}). Again, the expressions given here are general, and we need to invoke some PT or numerical treatments for a quantitative prediction of each term in Eq.~\eqref{eq:TKO_II_general}, which is left to future work.

\section{Explicit expressions for GI cross power spectra}
\label{sec:analytical_expressions}

In this section, we elaborate further upon the PT calculations to derive explicitly the analytical expressions for the PT model of the GI power spectrum. In doing so, we first need a prescription to relate, in real space, the galaxy density and ellipticity fields with the matter density field, $\delta$, namely, the galaxy bias relation. In general, these relations are nonlinear and also involve the stochastic contributions that are not solely characterized by the matter density field (see \cite{Desjacques_Jeong_Schmidt2018} for a comprehensive review). Implementing a general perturbative treatment of the galaxy density and shape bias developed recently, one can in principle compute the redshift-space power spectrum of tracer fields at given order. Although a complete and consistent PT calculation of the GI power spectrum is our final goal, we shall adopt here the simple linear bias relation between $\deltag$, $\gamma_{ij}$ and $\delta$ to conduct a proof-of-concept study and test the proposed PT modeling against cosmological $N$-body simulations. An extension of the linear relation to include nonlinear corrections to the galaxy density and shape bias as well as the stochasticities is straightforward (see the discussion in Sec.~\ref{sec:conclusion}), and we will address these issues in the forthcoming paper. 

Adopting the linear bias relation, the galaxy/halo density and ellipticity fields are expressed in Fourier space as
\begin{align}
\deltag (\bfk)&=b_1\,\delta(\bfk),
\label{eq:linear_bias}
\\
 \gamma_{ij}(\bfk) &=\bK\,\Bigl(\hat{k}_i\hat{k}_j-\frac{1}{3}\deltaK_{ij}\Bigr)\delta(\bfk),
\label{eq:LA_model}
\end{align}
where the coefficients $b_1$ and $\bK$ are called the linear galaxy/halo and linear shape bias parameters, respectively. The quantity $\hat{k}_i$ is the normalized wave vector given by $\hat{k}_i\equiv k_i/|\bfk_i|$. Note that the latter relation is referred to as the linear alignment model \cite{Catelan_etal2001,Hirata_Seljak2004}, and the coefficient $\bK$ typically takes negative value, $-1\lesssim\bK<0$.

Then, the real-space correlators involved in Eq.~\eqref{eq:TKO_GI_general} are expressed in term of the statistics of matter density and velocity-divergence fields:
\begin{align}
P_{{\rm g\gamma},ij}(\bfk)&=b_{\rm K}b_1 \Bigl(\hat{k}_i\hat{k}_j-\frac{1}{3}\deltaK_{ij}\Bigr)\,P_{\delta\delta}(k),
\label{eq:powerspec_g_gamma}
\\
P_{\theta \gamma,ij}(\bfk)&=b_{\rm K} \Bigl(\hat{k}_i\hat{k}_j-\frac{1}{3}\deltaK_{ij}\Bigr)\,P_{\theta\delta}(k),
\label{eq:powerspec_theta_gamma}
\\
\tilde{B}_{ij}(\bfk_1,\bfk_2,\bfk_3)&=b_{\rm K}
\Bigl(\hat{k}_{3,i}\hat{k}_{3,j}-\frac{1}{3}\deltaK_{ij}\Bigr) \Bigl\{ 
b_1\,B_{\theta\delta\delta}(\bfk_1,\bfk_2,\bfk_3) 
\nonumber
\\
&+ f\,\hat{k}_{2,z}^2 
\,B_{\theta\theta\delta}(\bfk_1,\bfk_2,\bfk_3)
\Bigr\},
\label{eq:bispec_bkij}
\end{align}
where the quantities $P_{\delta\delta}$, $P_{\theta\delta}$, and $B_{\theta X\delta}$ (X=$\delta$ or $\theta$) are respectively the autopower spectrum of matter density, cross power spectrum between velocity-divergence and matter density fields, and the cross bispectra among velocity-divergence and matter density fields. These are all defined in real space, similarly to the redshift-space counterparts at Eqs.~\eqref{eq:power_deltag_gamma}, \eqref{eq:power_theta_gamma} and \eqref{eq:cross_B_ij}.

Given the linear matter power spectrum, we can follow the same procedure of the standard PT calculations as used in the literature and compute the one-loop prediction for the real-space power spectra, $P_{\delta\delta}$ and $P_{\theta\delta}$, in a straightforward manner (e.g., Refs.~\cite{Nishimichi:2007xt,Taruya:2009ir}). For the cross bispectrum $B_{\theta X\delta}$ involved in the $A_{ij}$ term, the tree-level calculation suffices for a consistent one-loop prediction of the GI power spectrum [see Eqs.~\eqref{eq:tree_bispec} and \eqref{eq:2nd_PT_kernel} for its explicit expression].

On the other hand, in computing the $A_{ij}$ and $B_{ij}$ terms, the expressions given in Eqs.~\eqref{eq:A_ij} and \eqref{eq:B_ij} are not useful for an efficient PT computation. Because of the three-dimensional integral, the  tensor structure of these terms is not manifest, and it is not given in a factorizable form. Further, due to the RSD, the $A_{ij}$ and $B_{ij}$ terms are characterized not only by the wave number $k$ but also by the directional cosine $\mu$, but no explicit dependence on $\mu$ appears manifest. For an efficient PT calculation,  we follow and extend the technique used in Ref.~\cite{Taruya:2010mx} to derive the analytic expressions of  $A_{ij}$ and $B_{ij}$ terms, in which the tensor structure and the dependence of $\mu$ appear explicit and factorizable outside of the integrals, reducing also the dimensionality of the integral from 3D to 2D. 

In Appendix \ref{appendix:formulas_Aterm_Bterm}, we present perturbative expressions for the $A_{ij}$ and $B_{ij}$ terms valid at one-loop order. In practice, the actual observable of the galaxy/halo shape is the one projected onto the sky, and it would be rather useful to derive the expressions for the observed GI power spectrum. In doing so, it would be convenient to introduce the rotational-invariant ellipticity fields called $E$/$B$ modes \cite{Kamionkowski_etal1998,Crittenden_etal2002}. Writing the wave vector $\bfk$ as $\bfk=k(\sqrt{1-\mu^2}\cos\phi,\sqrt{1-\mu^2}\sin\phi,\mu)$, these are defined in Fourier space:\footnote{
References~\cite{Schmidt_Chisari_Dvorkin2015,Kurita_Takada2022} adopt a different convention for the projection operator at Eq.~\eqref{eq:pk_E/B-mode_projection}, which results  in a factor of $1/2$ difference in the overall amplitude at Eq.~\eqref{eq:TKO_GI_gE},   although this can be absorbed into the definition of the shape bias parameter, $b_{\rm K}$.
}
\begin{align}
\Bigl(
\begin{array}{c}
    \gamma_{\rm E}   \\
    \gamma_{\rm B}  
\end{array}
\Bigr)=\Bigl(
\begin{array}{cc}
    \cos(2\phi) &  \sin(2\phi)\\
    -\sin(2\phi) &  \cos(2\phi) 
\end{array}
\Bigr)\,\Bigl(
\begin{array}{c}
     \gamma_{xx}-\gamma_{yy}  \\
     2\gamma_{xy} 
\end{array}
\Bigr)
\label{eq:pk_E/B-mode_projection}
\end{align}
Correspondingly, the E-/B-mode decomposition of the GI cross power spectra can be made with $P_{{\rm g\gamma},ij}^{\rm(S)}$ through
\begin{align}
 \left(
\begin{array}{c}
P_{\rm gE}^{\rm(S)}(\bfk)  \\
P_{\rm gB}^{\rm(S)}(\bfk) 
\end{array}
\right) &=\left(
\begin{array}{cc}
 \cos(2\phi)& \sin(2\phi)\\
-\sin(2\phi)& \cos(2\phi)
\end{array}
\right)
\nonumber
\\
&\times 
\left(
\begin{array}{c}
P_{{\rm g\gamma},xx}^{\rm(S)}(\bfk)-P_{{\rm g\gamma},yy}^{\rm(S)}(\bfk)  \\
2P_{{\rm g\gamma},xy}^{\rm(S)}(\bfk) 
\end{array}
\right).
\label{eq:pk_E/B-mode_decomp}
\end{align}
Under the linear bias relations given at Eqs.~\eqref{eq:linear_bias} and \eqref{eq:LA_model}, we substitute Eq.~\eqref{eq:TKO_GI_general} into the above. With a help of Eqs.~\eqref{eq:powerspec_g_gamma}, \eqref{eq:powerspec_theta_gamma}, and the expressions of the $A_{ij}$ and $B_{ij}$ terms in Appendix \ref{appendix:formulas_Aterm_Bterm}, we find that the B-mode GI power spectrum becomes exactly vanishing, i.e., $P_{\rm gB}^{\rm(S)}=0$\footnote{This is solely due to the symmetric reason. That is, even if there exists a nonzero B-mode contribution in the ellipticity field, the B-mode GI spectrum becomes zero. Although the linear alignment model does not have nonzero B-mode contribution, it naturally arises when we go beyond the linear alignment model, leading to the nonvanishing auto B-mode power spectrum \cite{Kurita_etal2021,Bakx_etal2023,Georgiou_etal2024}. }. The nonzero contribution to the GI power spectrum comes only from the E-mode, and it is written as a function of the wavenumber $k$ and directional cosine $\mu$ in the following form:
\begin{align}
&    P_{\rm gE}^{\rm(S)}(k,\,\mu)  =D_{\rm FoG}(k\mu f\sigma_{\rm v})\,\bK\,(1-\mu^2)\,\Bigl[b_1\,P_{\delta\delta}(k) 
    \nonumber
    \\
    &\qquad + f\mu^2\,P_{\theta\delta}(k)+A_{\rm E}(k,\mu;\,b_1,\,f) + B_{\rm E}(k,\mu;\,b_1,\,f)\Bigr],
    \label{eq:TKO_GI_gE}
\end{align}
where $A_{\rm E}$ and $B_{\rm E}$ are derived from the $A_{ij}$ and $B_{ij}$ terms respectively, but the factor of $\bK(1-\mu^2)$ is removed, and is instead multiplied as an overall factor of the GI power spectrum. The explicit expressions of these terms, which are given in the IR-safe two-dimensional integral form, are summarized in Appendix \ref{appendix:formulas_AEterm_BEterm}.

Figure~\ref{fig:power_A_B_mu} plots each building block in the GI power spectrum, computed with the standard PT at the one-loop order. The results at $z=1$ are specifically shown, multiplying by the factor $k^{3/2}$. Here, the $A_{\rm E}$ and $B_{\rm E}$ terms are expanded in the polynomial form of $\mu$ and $f$ as $A_{\rm E}(k,\mu)=\sum_{n,m} f^m \,\mu^n\,A_n^m(k)$ and  $B_{\rm E}(k,\mu)=\sum_{n,m} f^m \mu^n\,B_n^m(k)$, respectively, and we plot their coefficients $A_n^m$ and $B_n^m$ respectively in left and right panels. Note that only the coefficients $A_2^1$, $B_2^2$ and $B_4^2$ are linearly dependent on the bias parameter $b_1$, and we set it to unity for simplicity. 
The coefficients of the $A_{\rm E}$ term exhibit wiggle feature arising from the BAOs and their amplitudes show an increase or decrease with the wave number. On the other hand, the coefficients of $B_{\rm E}$ term show a monotonic change in amplitude, although the variation of their amplitude is smaller than those of the $A_{\rm E}$. Figure~\ref{fig:power_A_B_mu} indicates that the correction terms not only change the broadband shape of the power spectrum, but also affect the BAOs mainly from the $A_{\rm E}$ term, giving additional smearing to the BAOs in redshift space.  

\begin{figure*}[tb]
 \includegraphics[width=12cm,angle=0]{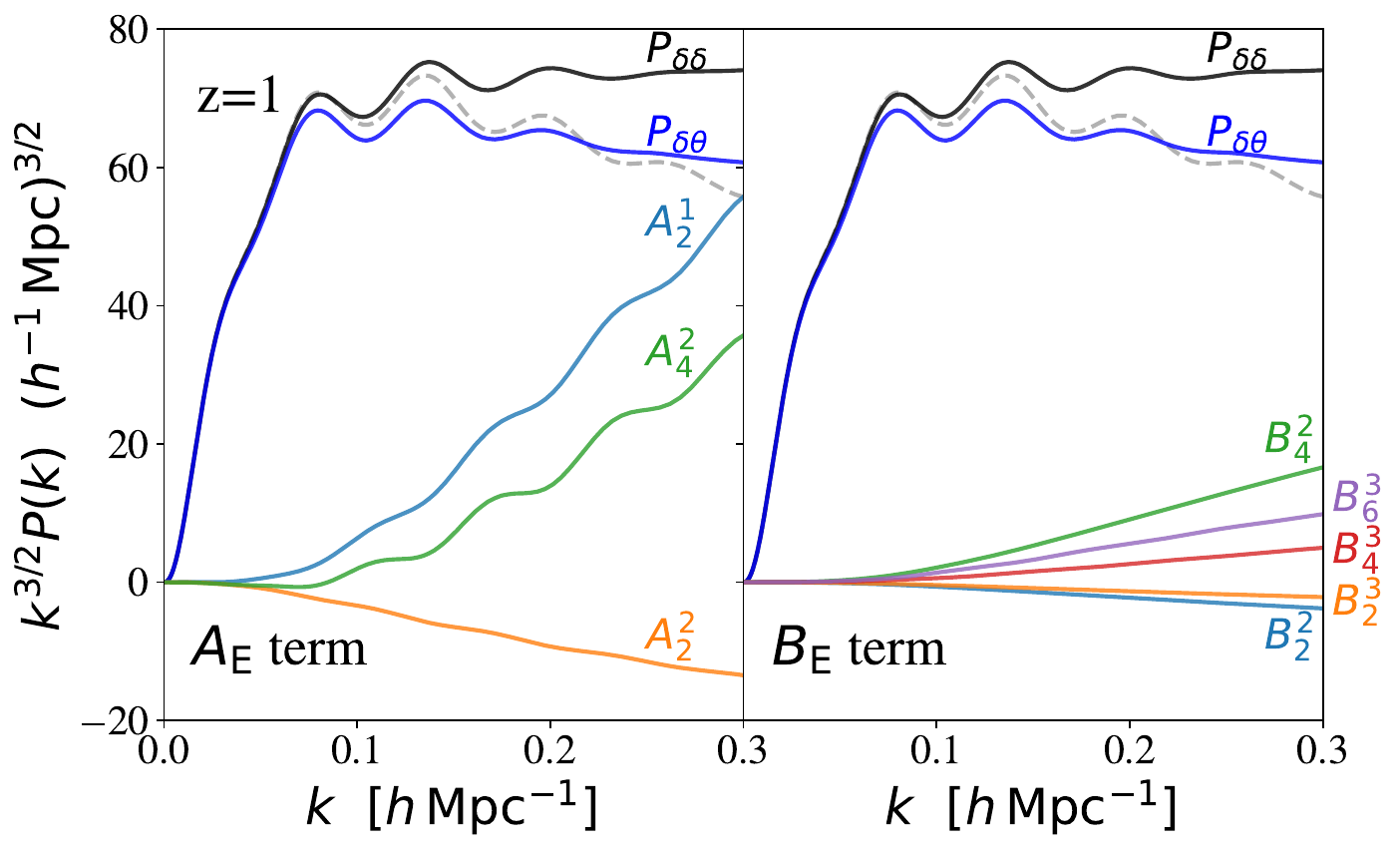}
\caption{Scale dependence of the building blocks in the GI power spectrum at Eq.~\eqref{eq:TKO_GI_gE}, obtained from the one-loop PT calculations. The results at $z=1$ are shown, and are compared with linear power spectrum depicted as gray dashed lines in both panels. Left panel: shows the real-space power spectra $P_{\delta\delta}$ and $P_{\delta\theta}$, and the $A_{\rm E}$ term, where the latter is expanded as $A_{\rm E}(k,\mu)=\sum_{n,m} f^m \,\mu^n\,A_n^m(k)$ and the coefficients $A_n^m$ are plotted. Right panel: plots the $B_{\rm E}$ term together with the  power spectra $P_{\delta\delta}$ and $P_{\delta\theta}$. Similarly, we expand it as $B_{\rm E}(k,\mu)=\sum_{n,m} f^m \mu^n\,B_n^m(k)$, and show the scale dependence of the coefficients $B_n^m$.  Note that while the coefficients $A_2^1$, $B_2^2$, and $B_4^2$ are linearly proportional to the linear bias parameter $b_1$, others are not. Here, we simply set $b_1$ to unity, and compare between the amplitudes of these coefficients. 
\label{fig:power_A_B_mu}
}
\end{figure*}
\begin{figure*}[tb]
 \includegraphics[width=15cm,angle=0]{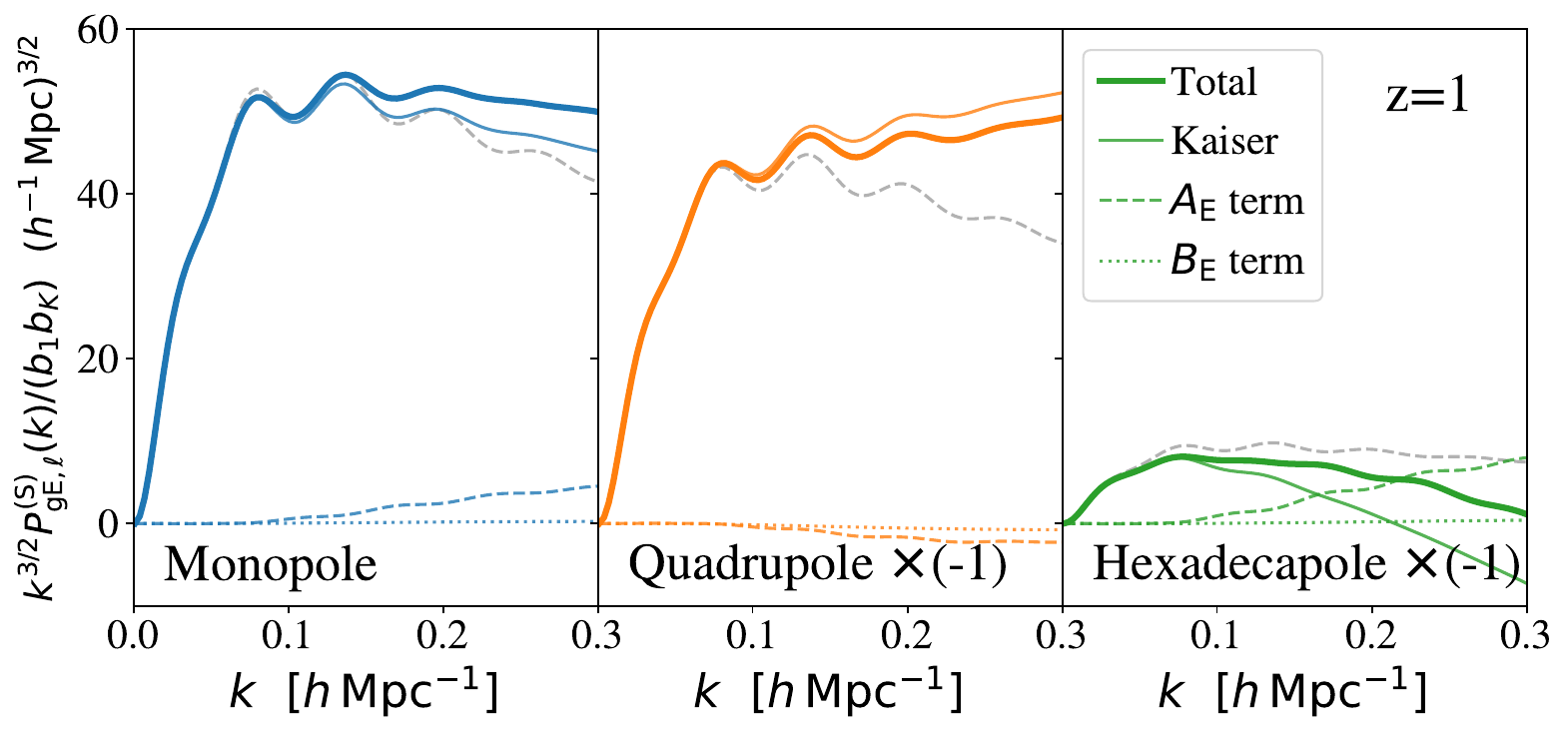}
\caption{Power spectrum multipoles $P^{\rm (S)}_{\rm gE,\ell}(\bfk)$ of the PT model at $z=1$. 
Convolving the building blocks shown in Fig.~\ref{fig:power_A_B_mu} with the Gaussian form of the damping function $D_{\rm FoG}$, the predicted GI cross spectra are plotted for multipole (left), quadrupole (middle) and hexadecapole (right) moments. Here, 
we set the bias parameters to $b_1=1.5$ and $b_{\rm K}=-0.2$, and 
the Gaussian damping function is evaluated with the linear theory estimate of the velocity dispersion (see the main text and footnote 8). In each panel, the total power spectrum, depicted as thick solid lines, is divided into three pieces, i.e., the contributions involving the Kaiser factor, the $A_{\rm E}$ and $B_{\rm E}$ terms, and these are respectively plotted as thin solid, dashed and dotted lines, together with linear theory predictions (gray dashed).   
\label{fig:power_A_B_multipoles}
}
\end{figure*}
In Fig.~\ref{fig:power_A_B_multipoles}, summing up all contributions at $z=1$ shown in Fig.~\ref{fig:power_A_B_mu}, we compute the GI cross power spectrum $P_{\rm gE}^{\rm(S)}$ in Eq.~\eqref{eq:TKO_GI_gE}, adopting $b_1=1.5$ and $\bK=-0.2$ as representative values of the bias parameters. We then plot the multipole power spectra, defined by
\begin{align}
P^{\rm(S)}_{{\rm g E},\ell}(k) =\frac{2\ell+1}{2}\int_{-1}^1d \mu\,P^{\rm(S)}_{\rm g E}(k,\mu) \mathcal{P}_\ell(\mu)
\label{eq:def_multipoles}
\end{align}
with $\mathcal{P}_\ell$ being the Legendre polynomials. Here and in what follows, we adopt the Gaussian form of the function $D_{\rm FoG}$, $D_{\rm FoG}(x)=\exp(-x^2)$. The relevance of its functional form has been long discussed in the literature, and we just follow the same form as frequently used in the autopower spectrum of galaxy/halo density fields, with $\sigma_{\rm v} $ being calculated from linear theory\footnote{In our definition, the $\sigma_{\rm v}^2$ is related to $\langle u_z^2\rangle$, and  (see Sec.~\ref{subsec:PT_model}) in linear theory, the prediction is given by
$\sigma_{\rm v}^2=\int dk\,P_{\rm lin}(k)/(6\pi^2)$. }. As we shall see in Sec.~\ref{sec:testing_model}, the predicted GI power spectrum with the Gaussian damping function can capture the nonlinear suppression of the power spectrum well, leading to a good agreement with simulations. In computing the angular integral in \eqref{eq:def_multipoles}, the following analytical formula is useful:
\begin{align}
& \int_{-1}^1 d\mu\,(1-\mu^2)\mu^n\,e^{-\alpha\mu^2}
\nonumber
\\
&\qquad=\frac{\gamma\bigl((n+1)/2,\,\alpha\bigr)}{\alpha^{(n+1)/2}}-\frac{\gamma\bigl((n+3)/2,\,\alpha\bigr)}{\alpha^{(n+3)/2}}
\\
&\qquad\simeq  \frac{4}{2n+3} \Bigl\{\frac{1}{2n+1}- \frac{\alpha}{2n+5}+\mathcal{O}(\alpha^2)\Bigr\}
\label{eq:integral_formula}
\end{align}
with the function $\gamma$ being the incomplete gamma function defined by $\gamma(s,\,x)=\int_0^x dt\,t^{s-1}e^{-t}$. Note that the second equality is valid for  $\alpha\ll1$.

In Fig.~\ref{fig:power_A_B_multipoles}, the total contributions to the multipole power spectra, depicted as thick solid lines, are divided into three pieces: the Kaiser term (solid) consisting of the density and velocity-divergence auto- and cross-power spectra, $A_{\rm E}$ term (dashed), and $B_{\rm E}$ term (dotted), all of which are convolved with the damping function $D_{\rm FoG}$. For reference, the linear theory predictions are also shown in gray dashed curves. As expected, the Kaiser term is the dominant contribution, but the contribution coming from the correction terms is not negligible, giving roughly a $10$\% change in amplitude. A closer look at their absolute values reveals that while the corrections enhance the power for the monopole and hexadecapole spectra, they suppress it for the quadrupole. Note that the same trends can be seen if we systematically increase the parameter $\sigma_{\rm v}$ \cite{Okumura_Taruya_Kurita_Nishimichi2024}, indicating that  with and without the correction terms, fitting the PT model to the observed or simulated spectra leads to different best-fit values for $\sigma_{\rm v}$. Since the linear growth rate $f$ also gives an overall change in the amplitude of power spectrum multipoles, this implies that if we allow $f$ to vary, the parameter estimation with the PT model may result in the biased $f$, depending on whether we include the correction terms or not. This point will be discussed carefully in the next section.

\section{Testing model predictions against simulations}
\label{sec:testing_model}

In this section, we are in position to discuss the validity and accuracy of the PT model predictions presented in Sec.~\ref{subsec:PT_model}. For this purpose, we used the $N$-body simulation data to measure the cross power spectrum between halo density and E-mode ellipticity fields. We then proceed to the parameter estimation study with the PT model template.

\subsection{Simulation data and GI cross spectra in real space}
\label{subsec:simulation}

The dataset of $N$-body simulations we used is a part of {\tt DarkQuest} \cite{Nishimichi_etal2019_DQ1}, consisting of 20 realization data of the box size $L_{\rm box}=1,000\,h^{-1}$\,Mpc and number of dark matter particles $N_{\rm dm}=2,048^3$, with output redshifts  $z=0.484$, $1.03$, and $1.48$. The cosmological parameters used in the simulations are those in the flat $\Lambda$CDM model, consistent with the Planck 2015 results \cite{Planck2015_XIII}\footnote{See [4] Planck TT,TE,EE+lowP of Table 3 in Ref~\cite{Planck2015_XIII}.}. We create the halo catalogs selected from the mass range, $[10^{12}, 10^{13}]\,h^{-1}{\rm M}_\odot$ with the halo finder, {\tt rockstar} \cite{Behroozi_etal2013}, including subhalo components. We then evaluate, for each halo, the shape inertia tensor, from which halo ellipticity fields, $\gamma_{\rm E}$ and $\gamma_{\rm B}$ are constructed, following Ref.~\cite{Kurita_etal2021}. In this paper, we use the matter, halo density and E-mode ellipticity fields to measure their auto- and cross-power spectra, adopting the same measurement method as in Ref.~\cite{Kurita_etal2021}. 

\begin{figure}[tb]
 \includegraphics[width=7.5cm,angle=0]{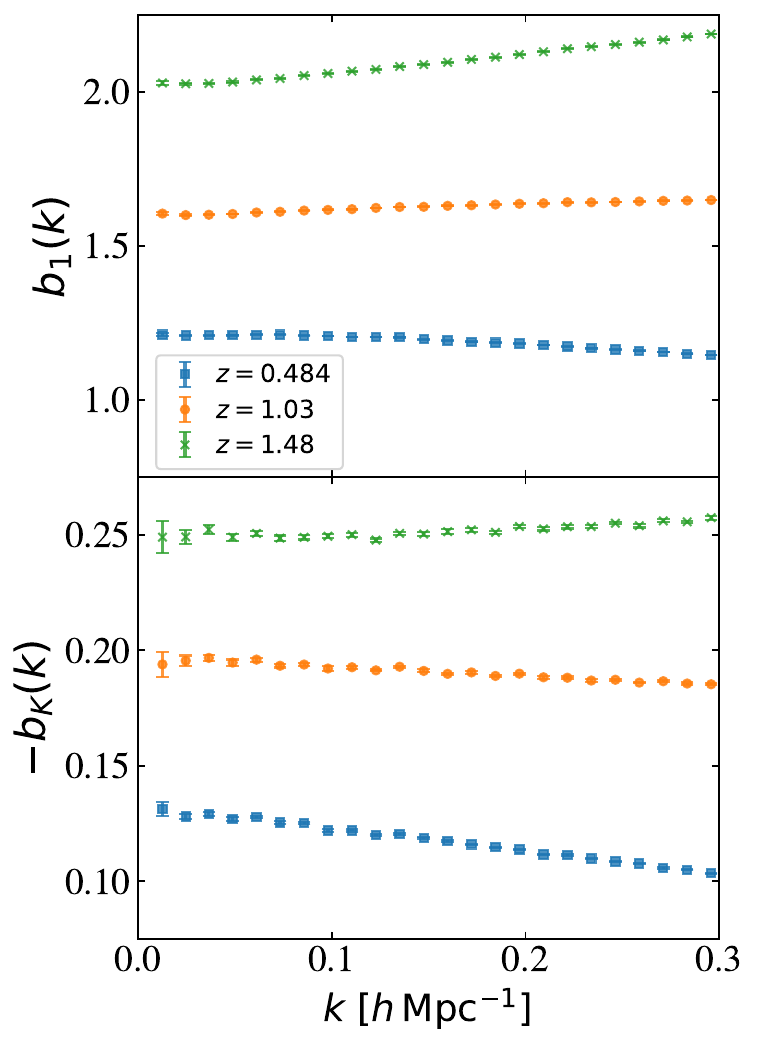}
\caption{Scale dependence of the halo density and shape bias, $b_1$ (upper) and $b_{\rm K}$ (lower), measured from simulations. In plotting the results, the latter is multiplied by $-1$. In both panels, crosses, filled circles, and filled squares are the results at $z=0.484$, $1.03$, and $1.48$, respectively. The linear density bias $b_1$ is estimated by taking the ratio of halo and matter density cross spectrum to the matter autopower spectra, i.e., $b_1=P_{\rm gm}(k)/P_{\rm mm}(k)$. On the other hand, the halo shape bias $b_{\rm K}$ is measured from the ratio of the monopole moment of the matter density and E-mode ellipticity cross spectra to the matter autopower spectra, further multiplied by the factor of $3/2$, i.e., $b_{\rm K}=(3/2)(P_{\rm mE}/P_{\rm mm})$.   
\label{fig:scale-dept_bias}
}
\end{figure}

In Fig.~\ref{fig:scale-dept_bias}, to see the properties of the halo catalogs, we plot the ratio of (monopole) real-space power spectra averaged over 20 realizations. To be precise, 
top and bottom 
panels respectively show $P_{\rm gm}/P_{\rm mm}$ and $(P_{\rm gE}/P_{\rm mm})$, the latter of which is further multiplied by the factor of $3/2$. Here, $P_{\rm gm}$, $P_{\rm mm}$ and $P_{\rm mE}$ are the halo and matter density cross power spectrum, matter density auto spectrum, and matter density and E-mode ellipticity cross spectrum, respectively.\footnote{For notational simplicity, we keep using the subscript $_{\rm g}$ even for the quantities related to halos.} Under the linear bias and linear alignment model in Eqs.~\eqref{eq:linear_bias} and \eqref{eq:LA_model}, the plotted results represent the linear bias and shape bias parameters, $b_1$ (upper) and $\bK$ (lower), respectively. 

Apart from the overall amplitude consistent with Ref.~\cite{Kurita_etal2021}\footnote{The ellipticity field defined in Ref.~\cite{Kurita_etal2021} is somewhat different from the one considered in this paper. While the halo shape measured from $N$-body simulations is directly linked to the ellipticity in our case, it is weighted by the so-called shear responsivity according to the weak lensing measurement \cite{Bernstein_Jarvis2002}. This leads to the different amplitude of $\bK$ by a factor of $2\mathcal{R}$, with the shear responsivity $\mathcal{R}$ being $\sim0.9$. 
Since the difference is absorbed into an overall scale-independent factor, it does not affect in general the cosmological parameter inference.}, Fig.~\ref{fig:scale-dept_bias} shows a mild scale-dependent feature in both the bias parameters, which becomes gradually prominent at $k\gtrsim0.1\,h$\,Mpc$^{-1}$. This indicates that the simple linear relation does not rigorously hold and higher-order bias terms/operators come to play a role \cite{Desjacques_Jeong_Schmidt2018,Bakx_etal2023}. Hence, going beyond the linear regime, accounting for the higher-order bias based on the effective field theory treatment (e.g., Refs.~\cite{2012JCAP...07..051B,2012JHEP...09..082C,Senatore_etal2015}, see also Ref.~\cite{Desjacques_Jeong_Schmidt2018}) would be important in properly modeling the statistics of biased tracers (see \cite{Vlah_Chisari_Schmidt2019,Vlah_Chisari_Schmidt2021,Bakx_etal2023} for the IA statistics in real space). 

Here, our primary focus is to test and validate the PT modeling in Sec.~\ref{subsec:PT_model}, and for this purpose, we stick to the analytical PT calculations based on the linear bias prescription in Sec.~\ref{sec:analytical_expressions}, leaving a consistent treatment with the effective field theory (EFT) approach to future work. To mitigate the impact of higher-order bias, we allow the linear bias parameters $b_1$ and $\bK$ to be scale dependent, and incorporate the measured results in Fig.~\ref{fig:scale-dept_bias} into the PT model predictions. 
\begin{figure*}[tb]
 \includegraphics[width=18cm,angle=0]{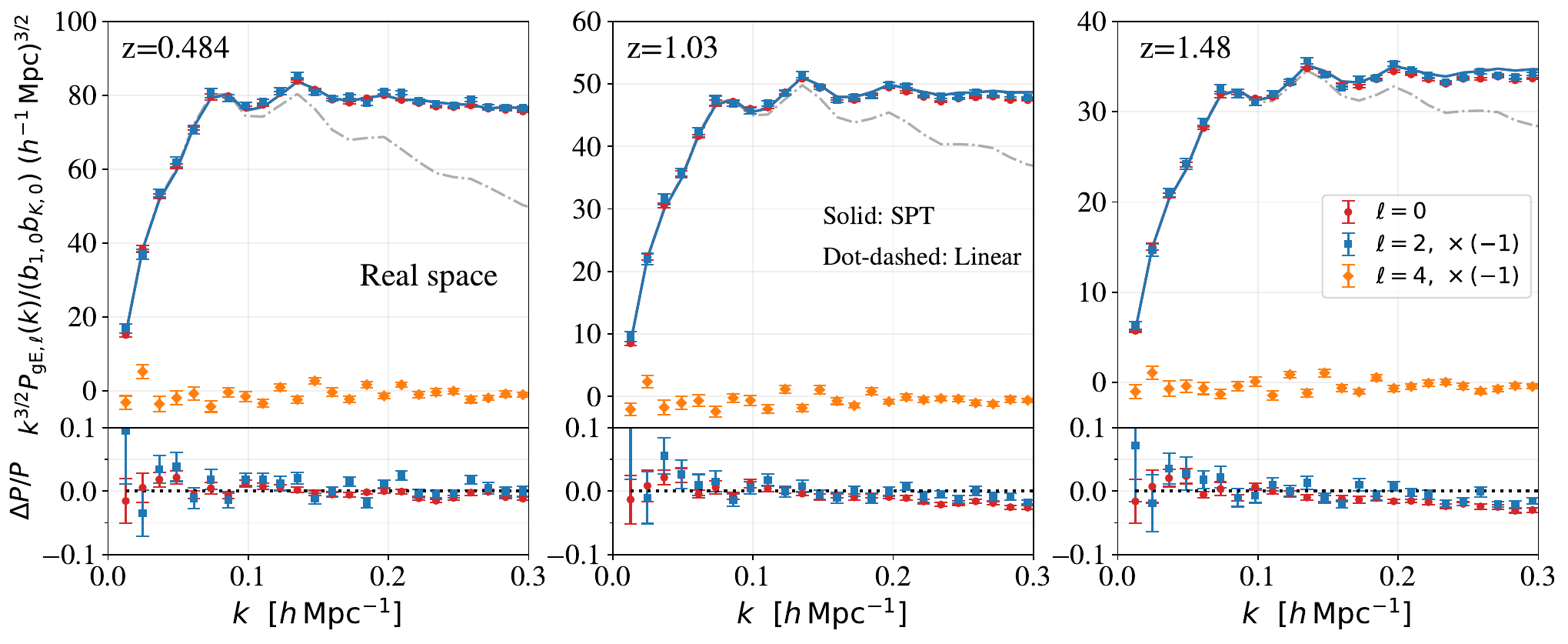}
\caption{Real-space cross power spectra, $P_{\rm gE,\ell}$ (upper panels). From left to right, the results at $z=0.484$, $1.03$, and $1.48$ are respectively shown. Here, the resultant power spectra are normalized by $b_{1,0}b_{\rm K,0}$, with $b_{1,0}$ and $b_{\rm K,0}$ being the low-$k$ limit of the bias parameters $b_1$ and $b_{\rm K}$ measured from $N$-body simulations (see Fig.~\ref{fig:scale-dept_bias}). Solid lines are PT predictions, and gray dot-dashed lines are the linear theory predictions. In both cases, we adopt the scale-dependent bias $b_1$ and $\bK$ measured in $N$-body simulations (see Fig.~\ref{fig:scale-dept_bias}). Note that in both the linear and PT predictions, the hexadecapole power spectra become vanishing. Also, the amplitudes of the monopole and quadrupole spectra are identical with each other, but with different signs. To see how the PT predictions resemble the simulated monopole and quadrupole, the fractional difference, $P^{\rm sim}_{\rm gE,\ell}(k)/P^{\rm PT}_{\rm gE,\ell}(k)-1$, is plotted for $\ell=0$ and $2$.
\label{fig:real-space_GI}
}
\end{figure*}
In Fig.~\ref{fig:real-space_GI}, to see if this treatment works well, we plot the multipole power spectra of the real-space GI correlation, i.e., $P_{\rm gE,\ell}$. The solid and dot-dashed lines are standard PT and linear theory predictions taking the scale-dependent linear bias into account. Starting from the linear matter power spectrum $P_{\rm lin}$ generated by the CMB Boltzmann solver {\tt class} \cite{CLASScode_2011}, these are obtained using the PT model prescription in Sec.~\ref{sec:analytical_expressions}, where the linear growth rate $f$ is set to zero and the damping function $D_{\rm FoG}$ is set to unity. Note that in linear theory, the power spectra $P_{\delta\delta}$ and $P_{\delta\theta}$ are identical and coincide with the linear power spectrum $P_{\rm lin}$. It is then shown in both the standard PT and linear theory calculations that the predicted monopole and quadrupole power spectra have identical amplitudes but opposite signs, while the hexadecapole power spectrum vanishes, i.e., $P_{\rm gE,0}=-P_{\rm gE,2}=(3/2)b_1\bK\,P_{\delta\delta}$ and $P_{\rm gE,4}=0$. 

Figure~\ref{fig:real-space_GI} shows that taking the scale-dependent bias into account, the standard PT predictions at one-loop order agree well with the monopole and quadrupole spectra measured from simulations over the plotted range, irrespective of redshifts. Also, the amplitude of hexadecapole spectrum in simulations is hardly distinguishable with zero, consistent with both the standard PT and linear theory predictions. 

In the lower panels of Fig.~\ref{fig:real-space_GI}, to closely look at the difference between PT predictions and simulations, we plot the fractional differences for monopole and quadrupole spectra, defined by $\Delta P/P\equiv P^{\rm sim}_{\rm gE,\ell}(k)/P^{\rm PT}_{\rm gE,\ell}(k)-1$, with $P^{\rm sim}_{\rm gE,\ell}$ and $P^{\rm PT}_{\rm gE,\ell}$ being respectively the power spectrum multipoles of the simulations and PT predictions. No notable difference is found between $\ell=0$ and $\ell=2$, although the simulation results at $k\gtrsim0.25\,h$\,Mpc$^{-1}$ exhibit a small departure from PT predictions, which looks relatively prominent at higher redshifts. We suspect that this is presumably due to the impact of nonlinear or higher-order bias, and hence the simple prescription with linear scale-dependent bias becomes inadequate. Nevertheless, the standard PT prediction reproduces the simulated GI spectrum quite well, covering a broader range than the prediction of the matter power spectrum $P_{\delta\delta}$, for which percent-level accuracy of the one-loop PT is restricted in the $\Lambda$CDM cosmology to the range $k\lesssim0.15\,h$\,Mpc$^{-1}$ to $k\lesssim0.1\,h$\,Mpc$^{-1}$ as we decrease the redshift from $1.5$ to $0.5$  (e.g., \cite{Carlson:2009it,Taruya:2009ir, Nishimichi:2008ry}). 

In any case, the results in Fig.~\ref{fig:real-space_GI} indicate that the prescription of the linear scale-dependent bias and linear alignment model work well. As long as we stick to the scales where the real-space prediction agrees well with simulations, we can go to redshift space and test the PT model in Sec.~\ref{subsec:PT_model}.

\subsection{Parameter inference}
\label{subsec:inference}

In this subsection, using the simulated power spectrum, we examine the parameter estimation study to test the redshift-space PT modeling. We are particularly interested in a robust determination of the growth of structure from the IA statistics.  Thus, adopting the scale-dependent linear bias $b_1$ and $\bK$ in Fig.~\ref{fig:scale-dept_bias} and using the analytical PT expressions in Sec~\ref{sec:analytical_expressions} as theoretical template, the parameters to be determined by the simulation data are the linear growth rate $f$ or $f\sigma_8$ and the parameter $\sigma_{\rm v}$ that controls the Fingers-of-God damping\footnote{In practice, the galaxy/halo density and shape bias have to be simultaneously determined on top of the parameters mentioned above. However, due to the degeneracy between $b_1$, $\bK$ and $f$, a meaningful parameter inference is made possible only when combining the GI cross spectrum with GG and II auto spectra, as demonstrated in Refs.~\cite{Taruya_Okumura2020,Okumura_Taruya2023}. Since our main objective is to present the PT model of the GI cross spectrum, we leave the practical parameter estimation to future work.}. 
\begin{table}[tb]
\caption{\label{tab:halo_num_shape} Number density and shape noise of the halo catalog evaluated from 20 realization data. }
\begin{ruledtabular}
\begin{tabular}{lcc}
Redshift & $n_{\rm g}$\, [$h^3$\,Mpc$^{-3}$] & $\sigma_\gamma$   
\\
\hline
0.484  & $4.18\times10^{-3}$ & $0.329$  
\\
1.03  & $3.67\times10^{-3}$ & $0.354$
\\
1.48 & $2.99\times10^{-3}$ & $0.368$
\end{tabular}
\end{ruledtabular}
\end{table}
\begin{figure*}[tb]
 \includegraphics[width=12cm,angle=0]{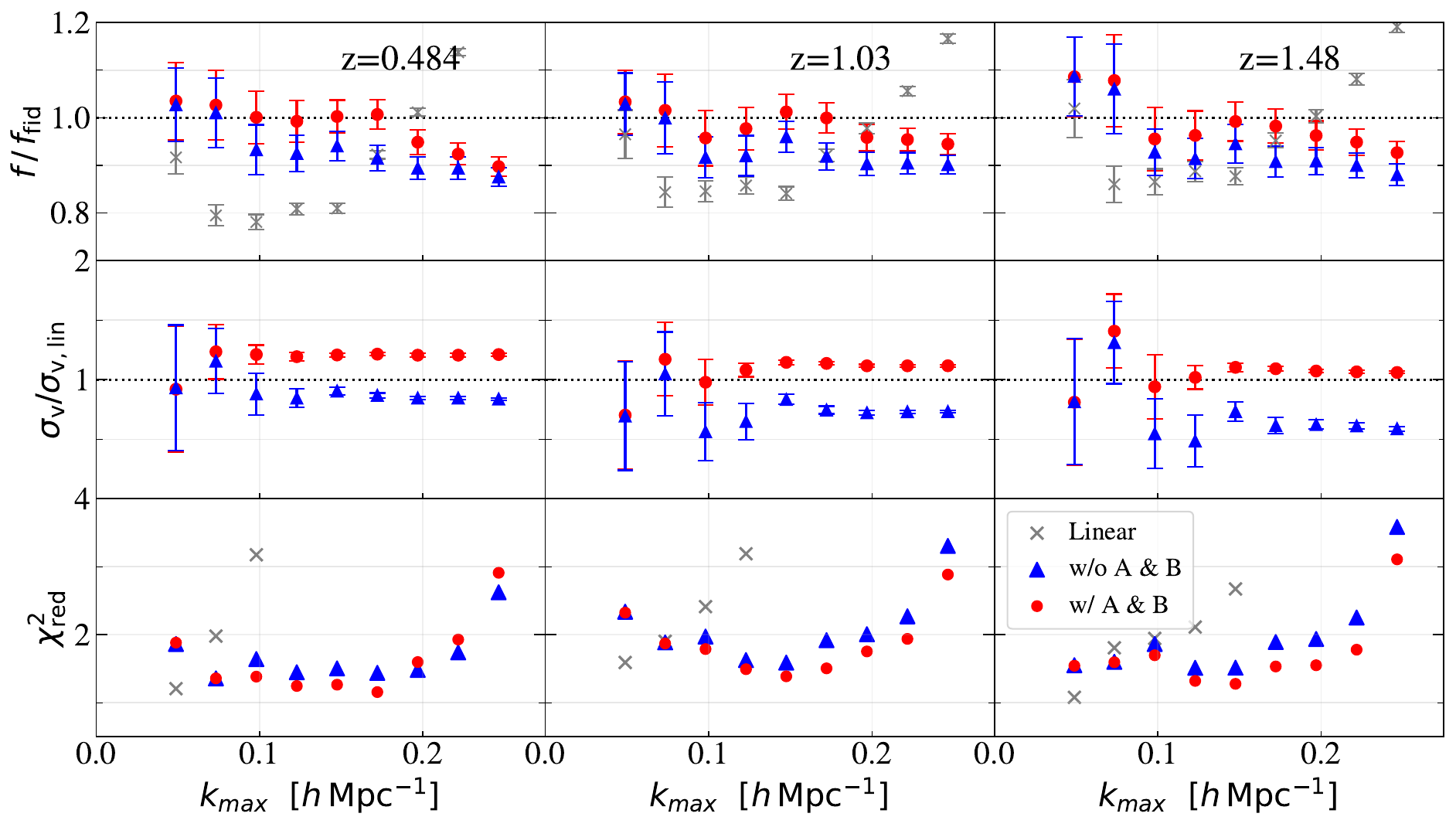}
\caption{One-dimensional marginalized errors on $f$ (upper) and $\sigma_{\rm v}$ (middle), and the reduced $\chi^2$ (lower), as a function of maximum wave number, $k_{\rm max}$. From left to right, the results at $z=0.484$, $1.03$, and $1.48$ are respectively shown. In each panel, both of the filled blue triangles and red circles represent the results based on the PT model at Eq.~\eqref{eq:TKO_GI_gE}, but for the latter case, the next-to-leading order corrections, $A_{\rm E}$ and $B_{\rm E}$, are ignored in the analysis of parameter inference. For reference, gray crosses are the linear theory template. In all cases, the linear scale dependent bias measured from simulations is incorporated into the predictions. 
\label{fig:1D_errors}
}
\end{figure*}

\begin{figure*}[tb]
 \includegraphics[width=18cm,angle=0]{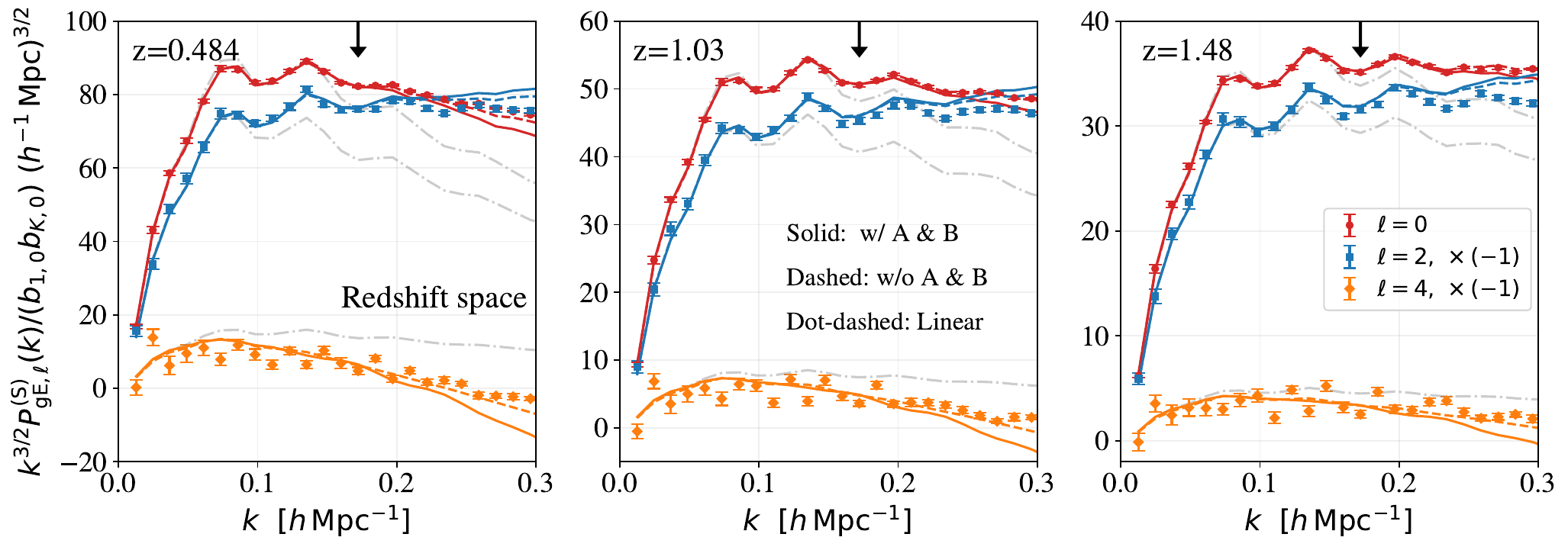}
\caption{Best-fit model of cross power spectra, $P_{\rm gE,\ell}^{\rm(S)}$ obtained by setting $k_{\rm max}$ to $0.172\,h$\,Mpc$^{-1}$. From left to right, the results at $z=0.484$, $1.03$, and $1.48$ are respectively shown. Here, the resultant power spectra are normalized by $b_{1,0}b_{\rm K,0}$, with $b_{1,0}$ and $b_{\rm K,0}$ being the low-$k$ limit of the bias parameters $b_1$ and $b_{\rm K}$ measured from $N$-body simulations. Note that the hexadecapole power spectra $(\ell=4)$ are not used for parameter inference, but are instead compared with the best-fit model predictions as a consistency check.
\label{fig:best-fit_GI}
}
\end{figure*}

We determine the best-fit values of these parameters by minimizing the following $\chi^2$, adopting the Gaussian power spectrum covariance, ${\rm Cov^{\rm G}}_{\ell,\ell'}$: 
\begin{align}
    \chi^2 & =\sum_{i=1}^{N_{\rm bin}}\sum_{\ell,\ell'} \Bigl[P^{\rm(S),sim}_{\rm gE,\ell}(k_i)-P^{\rm(S),PT}_{\rm gE,\ell}(k_i)\Bigr]
    \nonumber
    \\
&   \times \,\Bigl\{{\rm Cov^{\rm G}}_{\ell,\ell'}(k_i)\Bigr\}^{-1}\, \Bigl[P^{\rm(S),sim}_{\rm gE,\ell'}(k_i)-P^{\rm(S),PT}_{\rm gE,\ell'}(k_i)\Bigr]
\label{eq:chi2}
\end{align}
where, the spectra $P^{\rm(S),sim}_{\rm hE,\ell}$ and $P^{\rm(S),PT}_{\rm hE,\ell}$ are the measured power spectrum averaged over $20$ realizations and the PT model predictions, respectively. The quantity $N_{\rm bin}$ is the number of Fourier bins used in the analysis, which varies with $k_{\rm max}$. From Eq.~\eqref{eq:chi2}, the goodness of fit, $\chi^2_{\rm red}$, is also estimated with $\chi_{\rm red}^2\equiv\chi^2/N_{\rm dof}$, with $N_{\rm dof}$ being the number of degrees of freedom. For the two-parameter inference using the multipoles of $\ell=0$ and $2$, it is given by $N_{\rm dof}=2N_{\rm bin}-2$. It is known that hexadecapole power spectra is sensitive to the effects of finite grid size when measured from the grid-assigned density and ellipticity fields, leading to a noisy power spectrum shape \cite{Taruya:2013my}. This is partially seen in the real-space results in Fig.~\ref{fig:real-space_GI}. Since the signal-to-noise ratio of the $\ell=4$ data is lower than that of the $\ell=0$ and $2$ data and including it in the parameter inference would degrade the goodness of fit, 
we decided not to use it here\footnote{
Although it is somewhat computationally expensive, the effect of finite grid size can be incorporated into theoretical predictions, as demonstrated in Ref.~\cite{Taruya:2013my}. By applying the method proposed in Ref.~\cite{Taruya:2013my} (see their Appendix B), the hexadecapole power spectrum can be used for parameter inference, helping to reduce the statistical uncertainty of the best-fit parameters.}. 
Instead, we shall present the $\ell=4$ data for comparison with the best-fit results as a consistency check (see Fig.~\ref{fig:best-fit_GI}).

The covariance matrix ${\rm Cov}^{\rm G}_{\ell,\ell'}$ considered here assumes a diagonal form with respect to the wave number bins, but has off diagonal components among different multipoles, $\ell$ and $\ell'$. Taking the scale dependence of linear bias parameters $b_1$ and $b_{\rm K}$, we compute analytically the Gaussian covariance from linear theory. The analytical expression of the covariance is obtained from 
\begin{align}
&    {\rm Cov}^{\rm G}_{\ell,\ell'}(k)=\frac{(2\ell+1)(2\ell'+1)}{N_k}\int_{-1}^1\frac{d\mu}{2}\,\mathcal{P}_{\ell}^{\rm(S)}(\mu)\mathcal{P}_{\ell'}(\mu)
    \nonumber
    \\
&\quad \times    \Bigl[\Bigl\{P_{\rm gg}^{\rm(S)}(\bfk)+\frac{1}{n_{\rm g}}\Bigr\}\Bigl\{P_{\rm EE}^{\rm(S)}(\bfk)+\frac{\sigma_\gamma^2}{n_{\rm g}}\Bigr\}+\Bigl\{P_{\rm gE}^{\rm(S)}(\bfk)\Bigr\}^2\Bigr]
\label{eq:Gaussian_cov}
\end{align}
with the power spectra $P_{\rm XY}^{\rm(S)}$ (${\rm X, Y}=$ g or E) given by\footnote{At the linear order, the power spectrum $P^{\rm(S)}_{\rm EE}$ is identical to the one in the real space. }
\begin{align}
    P_{\rm XY}^{\rm(S)}(\bfk) =K_{\rm X}(\mu)K_{\rm Y}(\mu)\,P_{\rm lin}(k),\quad
    \left\{
    \begin{array}{l}
      K_{\rm g}=(b_1+\,f\,\mu^2)   \\
      \\
      K_{\rm E}=b_{\rm K}\,(1-\mu^2) 
    \end{array}
    \right.,
    \label{eq:linear_pk}
\end{align}
where the quantities $n_{\rm g}$ and  $\sigma_\gamma$ are the number density and shape noise of halos, respectively, both of which are obtained from the simulation dataset, summarized in Table \ref{tab:halo_num_shape}. The $N_{k}$ represents the number of Fourier modes for a given $k$, which is estimated from $N_k=4\pi\,k^2\,\Delta k/k_{\rm f}^3$, with $\Delta k$ and $k_{\rm f}$ being respectively the width of the Fourier bin and fundamental mode determined by the survey volume $V_{\rm s}$ through $k_{\rm f}\equiv 2\pi/V_{\rm s}^{1/3}$. In order to perform a stringent test, we here set the survey volume to the entire simulation volume over 20 realizations, i.e., $V_{\rm s}=20\,h^{-3}$\,Gpc$^3$. 
Note that the analytical formulas for the covariance above are partly given in Ref~\cite{Taruya_Akitsu2021} for the diagonal components of multipoles,  i.e., $\ell=\ell'$ (see Appendix G of their paper). In Appendix \ref{appendix:validity_covariance}, the validity of the Gaussian linear covariance is discussed, showing that the expression given at Eq.~\eqref{eq:Gaussian_cov} reasonably describes the diagonal components of the covariance measured from the $20$ realization data of $N$-body simulations (see Fig.~\ref{fig:cov}). 
\begin{figure*}[tb]
 \includegraphics[width=5.8
cm,angle=0]{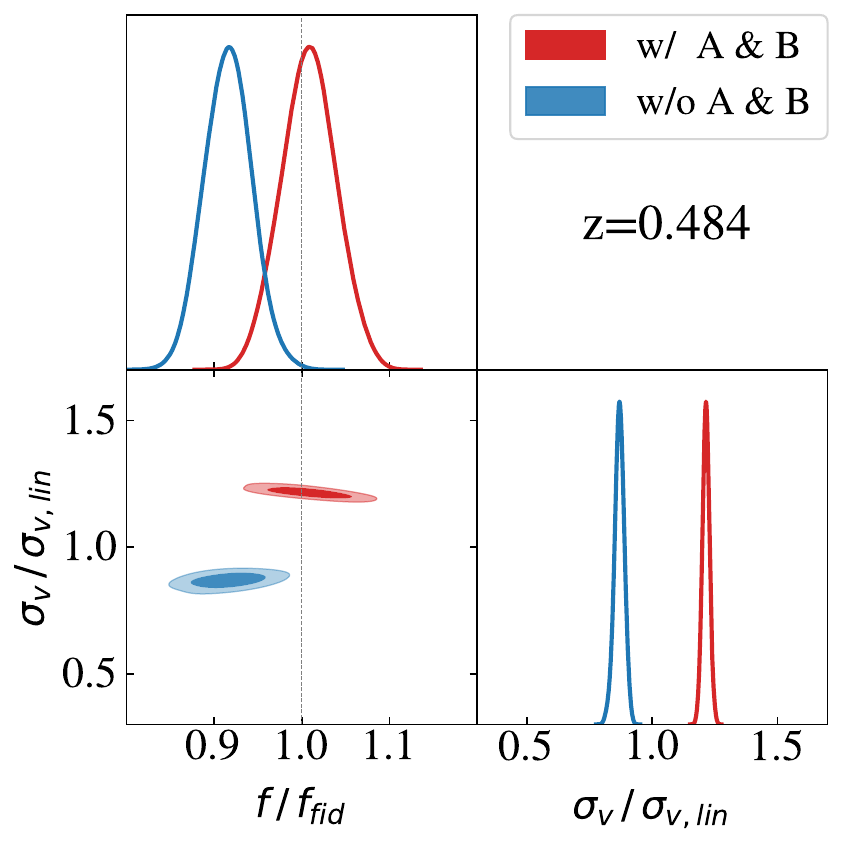}
\includegraphics[width=5.8
cm,angle=0]{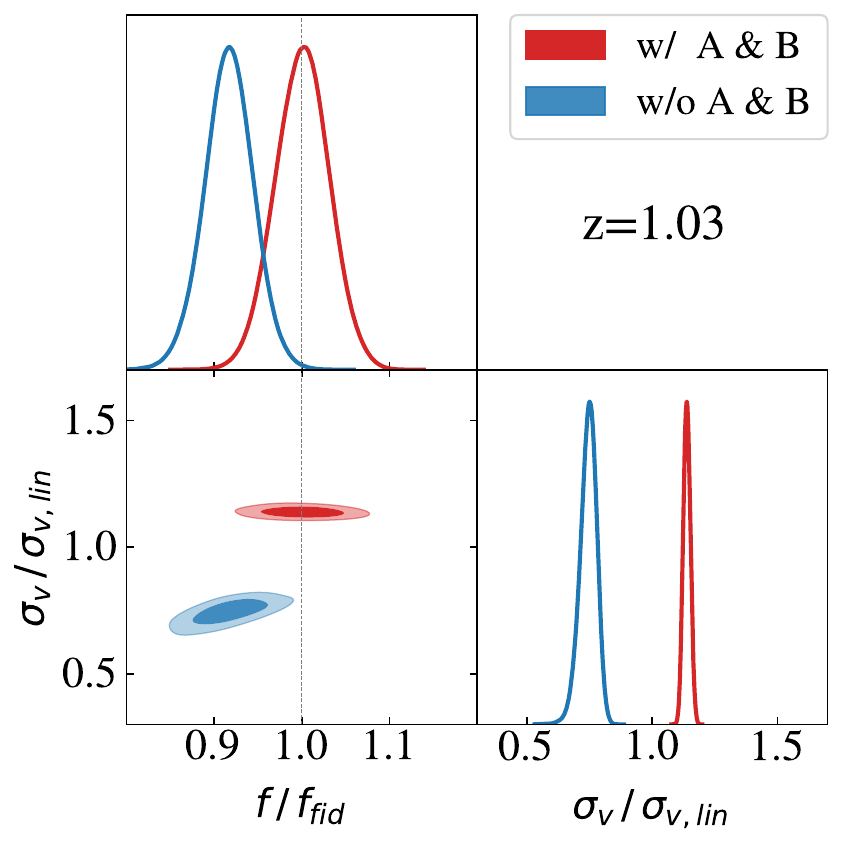}
\includegraphics[width=5.8
cm,angle=0]{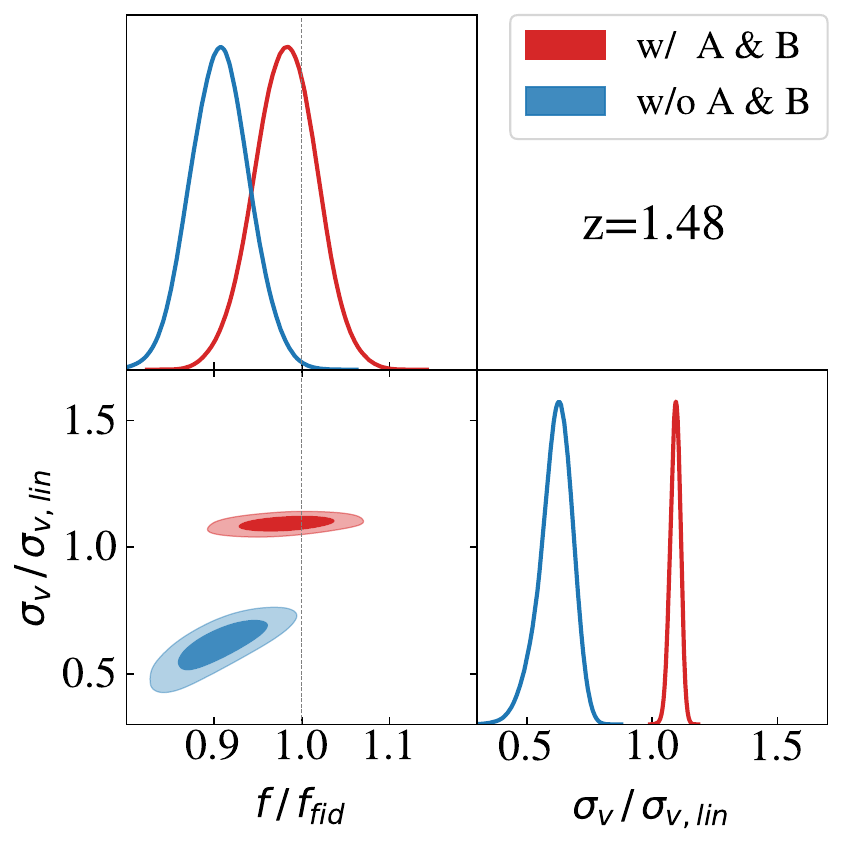}
\caption{Two-dimensional error contours on the parameters $f$ and $\sigma_{\rm v}$, normalized respectively by the fiducial value and linear theory prediction. From left to right, the results at $z=0.484$, $1.03$ and $1.48$ are respectively shown. In all panels, the maximum wave number $k_{\rm max}$ is set to $0.172\,h$\,Mpc$^{-1}$. 
\label{fig:2D_error_contour}
}
\end{figure*}

Adopting the sampling algorithm {\tt MultiNest} \cite{Feroz_Hobson2008,Feroz_Hobson_Bridges2009,Feroz_etal2019} with the wider flat prior range, $[0,2]$ for $f$ and $[0,15]$ $h\,$Mpc$^{-1}$ for $\sigma_{\rm v}$, the Bayesian parameter inference is performed, and the posterior distribution is obtained, varying the maximum wave number $k_{\rm max}$ at each redshift of $z=0.484$, $1.03$, and $1.48$. 

Figure~\ref{fig:1D_errors} summarizes the systematic survey of the parameter estimation, in which the one-dimensional marginalized errors on the linear growth rate and the FoG parameter are estimated from the posterior distribution and are respectively plotted around their mean values in upper and middle panels as a function of $k_{\rm max}$. Note that in all cases we examined, the mean values of the posterior distribution are  close enough to both the maximum likelihood and maximum a {\it posteriori} probability. Also, in the lower panel, the quantity $\chi_{\rm red}^2$, computed with the mean values of the estimated parameters, is plotted in logarithmic scales.

In Fig.~\ref{fig:1D_errors}  the results obtained from three different power spectrum templates are shown: (i) PT model given at Eq.~\eqref{eq:TKO_GI_gE}, including all terms (labeled as "w/ A \& B"), (ii) same model as in (i), but ignoring the correction terms, $A_{\rm E}$ and $B_{\rm E}$ (labeled as "w/o A \& B"), and (iii) linear theory template. Note that in (iii), the linear growth rate is the only parameter to estimate, and no results for $\sigma_{\rm v}$ are shown.

Clearly, even adopting the scale-dependent bias in simulations, the linear theory template fails to reproduce the fiducial value of $f$ beyond the linear scale. Although it accidentally matches the fiducial value at $k_{\rm max}\sim 0.2\,h$\,Mpc$^{-1}$, the goodness of fit, as indicated by $\chi_{\rm red}^2$, is rather poor, and falls outside the plot range (at $k_{\rm max}\sim 0.2\,h$\,Mpc$^{-1}$, $\chi^2_{\rm red}$ exceeds $100$, $35$, and $15$ at $z=0.484$, $1.03$, and $1.48$, respectively). The fitting result is thus considered to be unsuccessful. On the other hand, including the FoG damping and nonlinear corrections at one-loop order, both PT models, i.e., (i) and (ii), reproduce the simulation results quite well. This is irrespective of the redshift, having $\chi_{\rm red}^2\lesssim2$ even at $k_{\rm max}\sim0.2\,h$\,Mpc$^{-1}$. However,  differences are manifest for the estimated values of $f$ and $\sigma_{\rm v}$. Overall, in all of the redshifts examined, the PT model with the $A_{\rm E}$ and $B_{\rm E}$ terms tends to give a larger value of $f$ and $\sigma_{\rm v}$ than the model ignoring these corrections, and this results in reproducing well the fiducial values of the linear growth rate at $k_{\rm max}\lesssim0.18\,h$\,Mpc$^{-1}$. 
Given that the statistical error in this analysis is ideally small (dominated by the cosmic variance of hypothetical survey volume of $20\,h^{-3}$\,Gpc$^3$), this is a significant success.

To see the inference results in more detail, we pick up the result at $k_{\rm max}=0.172\,h$\,Mpc$^{-1}$, and the best-fit models of the power spectrum templates are compared with simulation results in Fig.~\ref{fig:best-fit_GI}. Also, in Fig~\ref{fig:2D_error_contour}, the estimated results of parameters $f$ and $\sigma_{\rm v}$ are plotted in two-dimensional contours, together with their one-dimensional marginalized posterior distributions. The best-fit PT models, both with and without the $A_{\rm E}$ and $B_{\rm E}$ terms, agree remarkably well with simulations for all multipoles, though we did not use the hexadecapole ($\ell=4)$ power spectra in the parameter inference. In both cases, a good agreement persists even beyond $k_{\rm max}$. However, as already shown in Fig.~\ref{fig:1D_errors}, the estimated parameter on $f$ is systematically biased if the correction terms are ignored in the PT model (blue contour). In two-dimensional error contours in Fig.~\ref{fig:2D_error_contour}, the bias is significant, and for all redshifts examined, the best-fitted linear growth rate is shifted away by more than $2\sigma$ from the fiducial value, showing also the degeneracy with FoG damping parameter $\sigma_{\rm v}$. Taking the $A_{\rm E}$ and $B_{\rm E}$ terms into account, the PT model reproduces the fiducial value fairly well within $1\sigma$ (red contour)\footnote{ The size of $1\sigma$ error ellipse changes with the dimension of parameter space (e.g., Chap. 11.5 of Ref.~\cite{Lupton_Book1993}). In two-dimensional parameter space, it becomes effectively larger than the one in the one-dimensional case by $2.3^{1/2}\simeq1.52$ for each axis if the posterior distribution is close to the Gaussian. }, having less degeneracy with $\sigma_{\rm v}$.

\section{Conclusion}
\label{sec:conclusion}

In this paper, we have presented a perturbation theory based model for the redshift-space power spectrum of galaxy/halo intrinsic alignments. As a tracer of the tidal force field of large-scale structure, galaxy/halo individual shapes or IAs can be a promising cosmological probe, and combining the shape/alignment information obtained from imaging surveys with the spectroscopic information, one can reconstruct, in three-dimensional space, the ellipticity fields defined as a symmetric trace-free tensor. The statistics of IAs therefore carry valuable cosmological information, complementary to conventional galaxy clustering statistics. In particular, the IA statistics have been shown to be sensitive to the redshift-space distortions and baryon acoustic oscillations, with these signals recently detected at high statistical significance. Hence, in combination with galaxy clustering data, future precision measurements of IAs from the so-called stage-IV galaxy surveys will play a crucial role to tighten cosmological constraints. One thus needs an accurate theoretical template for the IA statistics, especially beyond the linear theory predictions. 

The present paper provides one such template based on the PT. Despite its limitation, the PT based template has several advantages and offers a general framework to deal with various nonlinearities including gravitational evolution and galaxy density/shape bias. Along the line of this, we have presented an improved prescription of the redshift-space IA power spectra, capable of better controlling the Fingers-of-God damping (Sec.~\ref{subsec:PT_model}). While the power spectrum expressions we derived [Eqs.~\eqref{eq:TKO_GI_general} and \eqref{eq:TKO_II_general}] rely on the assumption that the spatial correlations of velocity fields are small and can be expanded perturbatively, they still preserve generality in the sense that building blocks in their expressions, consisting of the real-space correlators, can be computed with any PT treatment or numerical simulations. 

As a proof-of-concept study, we adopt the linear density and shape bias relations to further elaborate upon the PT calculations and derive the explicit PT expressions valid at one-loop order (Sec.~\ref{sec:analytical_expressions} and Appendixes~\ref{appendix:formulas_Aterm_Bterm} and \ref{appendix:formulas_AEterm_BEterm}). We then test our PT model of the GI cross power spectrum (to be precise, galaxy density and E-mode ellipticity cross power spectrum, $P_{\rm gE}^{\rm(S)}$) against the halo catalogs from cosmological $N$-body simulations. Taking properly the next-to-leading order corrections ($A_{\rm E}$ and $B_{\rm E}$) into account, we have shown that the present PT model provides an unbiased template, enabling us to measure the redshift-space distortions (i.e., linear growth rate) in a robust manner beyond the weakly nonlinear scales. 

Finally, the present work is considered as a first step toward proper PT modeling of the redshift-space IA statistics. As a consistent PT description, the EFT treatment would be essential to account for higher-order corrections to the galaxy density and shape bias \cite{Desjacques_Jeong_Schmidt2018,Vlah_Chisari_Schmidt2019,Vlah_Chisari_Schmidt2021,Bakx_etal2023}. To implement it consistently at one-loop order in our PT model of the redshift-space GI power spectrum $P^{\rm(S)}_{{\rm g\gamma},ij}$, one needs to update or elaborate the PT calculations for the three terms in Eq.~\eqref{eq:TKO_GI_general}: $P_{{\rm g\gamma},ij}$, $P_{\theta\gamma,ij}$, and $A_{ij}$, which respectively correspond to $P_{\rm gE}$, $P_{\rm \theta E}$, and $A_{\rm E}$ in the power spectrum $P^{\rm(S)}_{\rm gE}$ at \eqref{eq:TKO_GI_gE}\footnote{Adopting the linear density and shape bias relations in Eqs.~\eqref{eq:linear_bias} and \eqref{eq:LA_model}, the terms $P_{\rm gE}$ and $P_{\rm \theta E}$ are respectively replaced with $b_1\,P_{\delta\delta}$ and $\bK(1-\mu^2)\,P_{\theta\delta}$ in Eq.~\eqref{eq:TKO_GI_gE}.}. These are all real-space quantities, and the first two terms have been already discussed in Refs.~\cite{Desjacques_Jeong_Schmidt2018,Vlah_Chisari_Schmidt2019,Bakx_etal2023}, where relevant expressions at one-loop order are presented there. A nontrivial part may be the last term, $A_{ij}$ (or $A_{\rm E}$), which involves the cross bispectrum between density, velocity and ellipticity fields. However, the $A_{ij}$ term itself is intrinsically higher order, and the tree-level calculation of the bispectrum is sufficient for a consistent one-loop prediction of $P^{\rm(S)}_{{\rm g\gamma},ij}$, meaning that only the second-order EFT corrections come into play. In this respect, extending the present PT modeling to EFT descriptions would be straightforward. We will present it in a forthcoming paper, along with an explicit PT calculation of the II autopower spectrum.

\begin{acknowledgments}
We thank Takahiro Nishimichi for his useful comments and providing the $N$-body simulation data. We also thank Shogo Ishikawa for providing his simulation data which was helpful to validate the model prescription. 
This work was supported by MEXT/JSPS KAKENHI Grant Numbers JP20H05861 and JP21H01081 (AT). TO acknowledges support from the Ministry of Science and Technology of Taiwan under Grants No. NSTC 112-2112-M-001-034- and NSTC 113-2112-M-001-011-, and the Career Development Award, Academia Sinica (AS-CDA-108-M02) for the period of 2019-2023.
\end{acknowledgments}

\bigskip
\centerline{\bf Data Availability}

\smallskip
The data and analysis tools used to produce Figs.~\ref{fig:power_A_B_mu}--\ref{fig:cov} are openly available \cite{data_TKO}, the embargo period may apply.

\appendix
\section{Analytical expressions of $A_{ij}$ and $B_{ij}$ terms}
\label{appendix:formulas_Aterm_Bterm}

In this Appendix, adopting the linear bias relations in Eq.~\eqref{eq:linear_bias}, we present the analytical expressions of the next-to-leading order corrections, $A_{ij}$ and $B_{ij}$ terms, in the PT model at Eq.~\eqref{eq:TKO_GI_general}. Note that the tensor structure of the expressions given below is manifest, being constructed explicitly in terms of the wave vector $k_i$ and Kronecker delta $\deltaK_{ij}$. Hence it is given in a factorizable form outside of the integrals, reducing significantly the computational cost to obtain the $A_{ij}$ and $B_{ij}$ terms numerically.

\subsection{$A_{ij}$ term}
\label{subsec:formulas_A_ij}

Let us first write down the $A_{ij}$ term, given at Eq.~\eqref{eq:A_gx},  in the IR-safe integral form so that the integrand of this expression has the symmetry of $\bfp\leftrightarrow\bfk-\bfp$. This form allows us to avoid the apparent divergence around the pole in evaluating the integral over Fourier mode $\bfp$. We have 
\begin{align}
  A_{ij}(\bfk)&=\frac{k\mu\,f}{2}\int\frac{d^3\bfp}{(2\pi)^3}\,
\Biggl[\frac{p_z}{p^2}\Bigl\{\tilde{B}_{ij}(\bfp,\bfk-\bfp,-\bfk)
\nonumber
\\
&+\tilde{B}_{ij}(\bfp,-\bfk,\bfk-\bfp)\Bigr\}
+
\frac{k_z-p_z}{|\bfk-\bfp|^2}\Bigl\{\tilde{B}_{ij}(\bfk-\bfp,\bfp,-\bfk)
\nonumber
\\
&+\tilde{B}_{ij}(\bfk-\bfp,-\bfk,\bfp)\Bigr\}\Biggr].
\label{eq:Aij_term_IR-safe1}
\end{align}
Then, substituting the expression of $\tilde{B}_{ij}$ at Eq.~\eqref{eq:bispec_bkij} into the above,  the $A_{ij}$ term can be recast in the factorized form, in which the $\mu$ dependence is explicitly given as a series of polynomials
:  
\begin{align}
 A_{ij}(k,\mu)&=b_{\rm K}\,\sum_{n} \sum_{X=g,\theta} \mu^n\,\frac{k^3}{4\pi^2}\int_0^\infty dr \int_{-1}^1 dx\,
\nonumber
\\
& \times
\Bigl[ a_{{\rm X},ij}^{(n)} B_{\theta X\delta} (\bfp,\bfk-\bfp,-\bfk)
\nonumber
\\
&+ \tilde{a}_{{\rm X},ij}^{(n)} B_{\theta X\delta} (\bfp,-\bfk,\bfk-\bfp)
\nonumber
\\
&\qquad+  b_{{\rm X},ij}^{(n)} B_{\theta {\rm X}\delta} (\bfk-\bfp,\bfp,-\bfk)
\nonumber
\\
&+  \tilde{b}_{{\rm X},ij}^{(n)} B_{\theta {\rm X}\delta} (\bfk-\bfp,-\bfk,\bfp)
\Bigr].
\label{eq:Aij_term_IR-safe2}
\end{align}
where we introduced new variables $r$ and $x$ defined respectively by $r=p/k$ and $x=(\bfk\cdot\bfp)/(kp)$. The nonvanishing coefficients $a_{{\rm X},ij}^{(n)}$, $\tilde{a}_{{\rm X},ij}^{(n)}$, $b_{{\rm X},ij}^{(n)}$ and $\tilde{b}_{{\rm X},ij}^{(n)}$, given as functions of $r$ and $x$ with the factorizable tensor form outside of the integral, are summarized as follows:
\begin{align}
 a_{g,ij}^{(2)}&=f \frac{x}{2r}\,\Bigl(\hatk_i\hatk_j-\frac{1}{3}\deltaK_{i,j}\Bigr),
\label{eq:agij2}
\\
 a_{\theta,ij}^{(2)}&=f^2 \frac{(2-3 r x) (-1+x^2) }{4 (1+r^2-2 r x)}\,\Bigl(\hatk_i\hatk_j-\frac{1}{3}\deltaK_{i,j}\Bigr),
\label{eq:atij2}
\\
 a_{\theta,ij}^{(4)}&=f^2 \frac{2 x+r \{2-6 x^2+r x (-3+5 x^2)\}}{4 (1+r^2-2 r x)}
 \nonumber
 \\
 &\times\Bigl(\hatk_i\hatk_j-\frac{1}{3}\deltaK_{i,j}\Bigr),
\label{eq:atij4}
\end{align} 
for $a_{{\rm X},ij}^{(n)}$;
\begin{align}
 \tilde{a}_{{\rm g},ij}^{(1)}&=-f\,\frac{(-1+r x) (-1+x^2) }{4 (1+r^2-2 r x)}\,
\Bigl(\hatk_i\deltaK_{j,3}+\hatk_j\deltaK_{i,3}\Bigr),
\label{eq:tilde_agij1}
\\
 \tilde{a}_{{\rm g},ij}^{(2)}&=f\,\Biggl[\frac{ 2 x+r \{2-6 x^2+r x (-3+5 x^2)\}}{4 r (1+r^2-2 r x)}\,\hatk_i\hatk_j
 \nonumber
 \\
 &\,+\,\frac{x (-2+r (r+4 x-3 r x^2))}{12 r (1+r^2-2 r x)}\,\deltaK_{i,j}\Biggr],
\label{eq:tilde_agij2}
\\
 \tilde{a}_{\theta,ij}^{(3)}&=f\, \tilde{a}_{{\rm g},ij}^{(1)}\,
\label{eq:tilde_atij3}
\\
 \tilde{a}_{\theta,ij}^{(4)}&=f\, \tilde{a}_{{\rm g},ij}^{(2)}\,,
\label{eq:tilde_atij4}
\end{align}
for $\tilde{a}_{{\rm X},ij}^{(n)}$; 
\begin{align}
  b_{g,ij}^{(2)}&=f\, \frac{1-r x}{2(1+r^2-2rx)}\,\Bigl(\hatk_i\hatk_j-\frac{1}{3}\deltaK_{i,j}\Bigr),
\label{eq:bgij2}
\\
b_{\theta,ij}^{(2)}&=f^2\, \frac{(1-3 r x) (-1+x^2) }{4(1+r^2-2rx)}\,\Bigl(\hatk_i\hatk_j-\frac{1}{3}\deltaK_{i,j}\Bigr),
\label{eq:btij2}
\\
b_{\theta,ij}^{(4)}&=f^2\, \frac{-1+3 x^2+r x (3-5 x^2) }{4(1+r^2-2rx)}\,\Bigl(\hatk_i\hatk_j-\frac{1}{3}\deltaK_{i,j}\Bigr),
\label{eq:btij4}
\end{align}
for $b_{{\rm X},ij}^{(n)}$; and
\begin{align}
 \tilde{b}_{{\rm g},ij}^{(1)}&=f\,\frac{rx(-1+x^2)}{4(1+r^2-2rx)}\,\Bigl(\hatk_i\deltaK_{j,3}+\hatk_j\deltaK_{i,3}\Bigr),
\label{eq:tilde_bgij1}
\\
 \tilde{b}_{{\rm g},ij}^{(2)}&=f\,\Biggl[\frac{-1+3 x^2+r x (3-5 x^2)}{4(1+r^2-2rx)}\,\hatk_i\hatk_j
 \nonumber
 \\
 &+\frac{(-1+r x) (-1+3 x^2)}{12(1+r^2-2rx)}\,\deltaK_{i,j}
\Biggr],
\label{eq:tilde_bgij2}
\\
\tilde{b}_{\theta,ij}^{(3)}&=f\,\tilde{b}_{{\rm g},ij}^{(1)},
\label{eq:tilde_btij3}
\\
\tilde{b}_{\theta,ij}^{(4)}&=f\,\tilde{b}_{{\rm g},ij}^{(2)},
\label{eq:tilde_btij4}
\end{align}
for $\tilde{b}_{{\rm X},ij}^{(n)}$. Note that the actual IR-safe integration of the $A_{ij}$ term can be performed by replacing the integral in Eq.~\eqref{eq:Aij_term_IR-safe2} with the one given below\footnote{This also applies to the integral in Eqs.~\eqref{eq:Bij_term}, \eqref{eq:AE_term_IR-safe}, and \eqref{eq:BE_term}.}: 
\begin{align}
 \int_0^\infty dr\int_{-1}^1 dx \longrightarrow  2\,\int_0^\infty dr\int_{-1}^{1/(2r)} dx.
\end{align}

In deriving the expressions of the coefficients above, we used the following formulas to factorize the directional dependence ($\mu$) and vectorial components [assuming that the function $f(\bfk,\bfp)$ depends only on $k$, $r$ and $x$]:
\begin{widetext}
\begin{align}
 \int \frac{d^3\bfp}{(2\pi)^3}\,p_i\, f(\bfk,\bfp)&=k^i\, \frac{k^3}{(4\pi^2)}
\int_0^\infty dr \int_{-1}^1 dx\,rx\,f(\bfk,\bfp),
\\
 \int \frac{d^3\bfp}{(2\pi)^3}\,p_ip_j\, f(\bfk,\bfp)&=
\frac{k^5}{(4\pi^2)} \int_0^\infty dr \int_{-1}^1 dx\,\frac{r^2}{2}\Bigl\{(1-x^2)\,\deltaK_{ij} -(1-3x^2)\,\hatk_i\hatk_j\Bigr\}\,f(\bfk,\bfp),
\\
 \int \frac{d^3\bfp}{(2\pi)^3}\,p_ip_jp_k\, f(\bfk,\bfp)&=\,
\frac{k^6}{(4\pi^2)} \int_0^\infty dr \int_{-1}^1 dx\,\frac{r^3}{2}\,\Bigl\{x(1-x^2)\,\bigl(\deltaK_{ij}\hatk_k+\deltaK_{ik}\hatk_j+\deltaK_{jk}\hatk_i\bigr)
+ 
(-3x+5x^3)\,\hatk_i\hatk_j\hatk_k\Bigr\}\,f(\bfk,\bfp).
\end{align}
\end{widetext}

Note that the traceless condition ${\rm Tr}[A_{ij}]=0$ 
consistently holds for the expressions presented above as follows. For $a_{{\rm X},ij}^{(n)}$ and $b_{{\rm X},ij}^{(n)}$, each of the coefficients individually satisfies the traceless condition, i.e., ${\rm Tr}[ a_{{\rm X},ij}^{(n)}] =0$ and ${\rm Tr}[ b_{{\rm X},ij}^{(n)}] =0$. On the other hand, for the coefficients $\tilde{a}_{{\rm X},ij}^{(n)}$ and $\tilde{b}_{{\rm X},ij}^{(n)}$, 
the following relation is shown to be satisfied:
\begin{align}
 {\rm Tr}[ \tilde{a}_{{\rm g},ij}^{(1)}\,\mu+\tilde{a}_{{\rm g},ij}^{(2)}\mu^2] &=0=f\, {\rm Tr}[ \tilde{a}_{\theta,ij}^{(3)}\,\mu^3+\tilde{a}_{\theta,ij}^{(4)}\mu^4],
 \\
{\rm Tr}[ \tilde{b}_{{\rm g},ij}^{(1)}\,\mu+\tilde{b}_{{\rm g},ij}^{(2)}\mu^2] &=0=f\, {\rm Tr}[ \tilde{b}_{\theta,ij}^{(3)}\,\mu^3+\tilde{b}_{\theta,ij}^{(4)}\mu^4]. 
\end{align}
Thus, combining these relations, the $A_{ij}$ term is shown to be trace free. 

Finally, for more quantitative calculation of the $A_{ij}$ term, one needs the explicit expression for the cross bispectrum $B_{\theta{\rm X}\delta}$ $(X=\delta,\,\theta)$. For a consistent one-loop calculation of the GI cross power spectrum, the tree-level bispectrum suffices, which is explicitly given below:
\begin{align}
    B_{\theta{\rm X}\delta}(\bfk_1,\bfk_2,\bfk_3)&=\,2\,\Bigl\{\mathcal{F}_\theta^{(2)}(\bfk_2,\bfk_3)P_{\rm lin}(k_2)P_{\rm lin}(k_3)
    \nonumber
    \\
    &+ \mathcal{F}_{\rm X}^{(2)}(\bfk_3,\bfk_1)P_{\rm lin}(k_3)P_{\rm lin}(k_1)
    \nonumber
    \\
    & + \mathcal{F}_\delta^{(2)}(\bfk_1,\bfk_2)P_{\rm lin}(k_1)P_{\rm lin}(k_2)\Bigr\}, 
    \label{eq:tree_bispec}
\end{align}
where $P_{\rm lin}$ is the linear matter power spectrum. The function $\mathcal{F}_{\rm X}^{(2)}$ $(X=\delta,\,\theta)$ is the second-order standard PT kernel defined by
\begin{align}
    \mathcal{F}_{\rm X}^{(2)}(\bfk_1,\bfk_2)&=\frac{a_{\rm X}}{7}+\frac{1}{2}\Bigl(\frac{\bfk_1\cdot\bfk_2}{k_1k_2}\Bigr)\Bigl(\frac{k_1}{k_2}+\frac{k_2}{k_1}\Bigr)
    \nonumber
    \\
    &+\frac{b_{\rm X}}{7}\Bigl(\frac{\bfk_1\cdot\bfk_2}{k_1k_2}\Bigr)^2
    \label{eq:2nd_PT_kernel}
\end{align}
with the coefficients $a_{\rm X}$ and $b_{\rm X}$ being $(a_{\rm X}, b_{\rm X})=(5,2)$ for ${\rm X}=\delta$ and $(3,4)$ for ${\rm X}=\theta$.


\subsection{$B_{ij}$ term}
\label{subsec:formulas_B_ij}

Similarly, the $B_{ij}$ term of the GI cross power spectra, defined in Eq.~(\ref{eq:B_ij}), is rewritten with the IR-safe integral form:
\begin{align}
 B_{ij}(\bfk)&= \frac{(k\,\mu f)^2}{2} \int \frac{d^3\bfp}{(2\pi)^3}\,\Bigl\{
F(\bfp)G_{ij}(\bfk-\bfp) 
\nonumber
\\
&+ F(\bfk-\bfp)G_{ij}(\bfp)\Bigr\};
\\
&
F(\bfp)=\frac{p_z}{p^2}\Bigl\{P_{g\theta}(p)+f\,\frac{p_z^2}{p^2} \,P_{\theta\theta}(p)\Bigr\},
\nonumber
\\
&G_{ij}(\bfp)=\frac{p_z}{p^2}\,\bigl(\hatp_i\hatp_j-\frac{1}{3}\deltaK_{ij}\bigr)P_{\theta \delta}(p).
\end{align}
Based on this form, the expressions in which the $\mu$ dependence and the tensor structure are explicitly given outside of the integral are derived, reducing also the dimensionality of the integral. We obtain
\begin{align}
 B_{ij}(k,\mu)&=b_{\rm K}\,\sum_{n}  \mu^n\,\frac{k^3}{4\pi^2}\,\int_0^\infty dr\int_{-1}^1 dx\,
\nonumber
\\
&\times
\Biggl[\,\Bigl\{f^2\, \alpha_{ij}^{(n)} P_{{\rm g}\theta} (k r) 
+ f^3\,\beta_{ij}^{(n)} P_{\theta\theta} (k r) 
\Bigr\}\,
\nonumber
\\
&\times \frac{P_{\theta\delta}(k\sqrt{1+r^2-2rx})}{(1+r^2-2rx)^2}
\nonumber
\\
&
+\Bigl\{f^2\, \tilde{\alpha}_{ij}^{(n)} P_{{\rm g}\theta} (k\sqrt{1+r^2-2rx} ) 
\nonumber
\\
&
+ f^3\,\tilde{\beta}_{ij}^{(n)} P_{\theta\theta} (kk\sqrt{1+r^2-2rx}) 
\Bigr\}\,
\nonumber
\\
&
\times\frac{P_{\theta\delta}(k\,r)}{(1+r^2-2rx)^2}
\Biggr]
\label{eq:Bij_term}
\end{align}
with the nonvanishing coefficients $\alpha_{ij}^{(n)}$, $\beta_{ij}^{(n)}$, 
$\tilde{\alpha}_{ij}^{(n)}$ and $\tilde{\beta}_{ij}^{(n)}$ summarized as follows:
\begin{widetext}
\begin{align}
\alpha_{ij}^{(2)} & = \,(-1+x^2)\Bigl[ \frac{1}{16}\bigl\{4-8 r x+r^2 (-1+5 x^2)\bigr\}\, \hatk_i\hatk_j 
-\frac{1}{8} r^2 (-1+x^2) \,\deltaK_{i,3}\deltaK_{j,3} -\frac{1}{48}\,\bigl\{4+r (r-8 x+3 r x^2)\bigr\}\,\deltaK_{i,j}\Bigr],
\label{eq:alpha_ij2}
\\
\alpha_{ij}^{(3)} & = \frac{1}{8}\,(-1+x^2) \bigl\{2-6 r x+r^2 (-1+5 x^2)\bigr\} \Bigl(\hatk_i\deltaK_{j,3}+\hatk_j\deltaK_{i,3}\Bigr),
\label{eq:alpha_ij3}
\\
\alpha_{ij}^{(4)} & = -\frac{1}{16r} \bigl[-8 x+r \bigl\{-12+36 x^2+12 r x (3-5 x^2)+r^2 (3-30 x^2+35 x^4)\bigr\}\bigr] \,\hatk_i\hatk_j
\nonumber
\\
& +\frac{1}{48r} \bigl[-8 x+r \bigl\{-4+28 x^2+12 r x (1-3 x^2)+r^2 (-1-6 x^2+15 x^4)\bigr\}\bigr]\,\deltaK_{i,j}
\label{eq:alpha_ij4}
\end{align}
for $\alpha_{ij}^{(n)}$;
\begin{align}
 \beta_{ij}^{(2)} & = (-1+x^2)^2 \Bigl[\frac{1}{32}\bigl\{-6+12 r x+r^2 (1-7 x^2)\bigr\}\,\hatk_i\hatk_j + \frac{1}{8}\,r^2(-1+x^2)\,\deltaK_{i,3}\deltaK_{j,3} 
\nonumber
\\
&+ \frac{1}{32}\bigl\{2-4 r x+r^2 (1+x^2)\bigr\}\,\deltaK_{i,j}\Bigr],
\label{eq:beta_ij2}
\\
 \beta_{ij}^{(3)} & = \frac{1}{16}\,(-1+x^2)^2 \bigl\{3-15 r x+2 r^2 (-1+7 x^2)\bigr\}\,\Bigl(\hatk_i\deltaK_{j,3}+\hatk_j\deltaK_{i,3}\Bigr),
\nonumber
\\
 \beta_{ij}^{(4)} & = \frac{1}{16\,r}(-1+x^2)\,\Bigl[3\bigl\{-4 x+4 r (-1+5 x^2)-5 r^2 x (-3+7 x^2)+r^3 (1-14 x^2+21 x^4)\bigr\}\,\hatk_i\hatk_j
\nonumber
\\
& + \bigl\{-2 r^2 (-1+x^2) (-3 x+r (-1+7 x^2)\bigr\}\,\deltaK_{i,3}\deltaK_{j,3}
\nonumber
\\
& + \bigl\{4 x+r (2-18 x^2)+r^2 x (-3+23 x^2)+r^3 (1-2 x^2-7 x^4)\bigr\}\,\deltaK_{i,j}\Bigr],
\label{eq:beta_ij4}
\\
 \beta_{ij}^{(5)} & = \frac{1}{16}\,(-1+x^2)\bigl\{-3+15 x^2-7 r x (-3+7 x^2)+r^2 (2-28 x^2+42 x^4)\bigr\}\,\Bigl(\hatk_i\deltaK_{j,3}+\hatk_j\deltaK_{i,3}\Bigr), 
\label{eq:beta_ij5}
\\
 \beta_{ij}^{(6)} & = -\frac{1}{32\,r}\,
\bigl\{8 x (3-5 x^2)+6 r (3-30 x^2+35 x^4)-6 r^2 x (15-70 x^2+63 x^4)
\nonumber
\\
&+r^3 (-5+105 x^2-315 x^4+231 x^6)\bigr\}\,\hatk_i\hatk_j
\nonumber
\\
&+ \frac{1}{96\,r}\,\bigl\{8 x (3-5 x^2)+6 r (1-18 x^2+25 x^4)-2 r^2 x (3-70 x^2+91 x^4)
\nonumber
\\
&+r^3 (3-15 x^2-35 x^4+63 x^6)\bigr\}\,\deltaK_{i,j},
\label{eq:beta_ij6}
\end{align}
for $\beta_{ij}^{(n)}$;
\begin{align}
 \tilde{\alpha}_{ij}^{(2)} & = -(1+r^2-2 r x) (-1+x^2)\Bigl\{\frac{1}{16}\, (1-5x^2)\,\hatk_i\hatk_j+\frac{1}{8}(-1+x^2)\,\deltaK_{i,3}\deltaK_{j,3} + \frac{1}{48}\,(1+3x^2)\,\deltaK_{i,j}\Bigr\},
\label{eq:tilde_alpha_ij2}
\\
 \tilde{\alpha}_{ij}^{(3)} & = \frac{1}{8\,r}\,(1+r^2-2 r x) (-1+x^2) \bigl\{-2 x+r (-1+5 x^2)\bigr\}\,\Bigl(\hatk_i\deltaK_{j,3}+\hatk_j\deltaK_{i,3}\Bigr),
\label{eq:tilde_alpha_ij3}
\\
 \tilde{\alpha}_{ij}^{(4)} & = -(1+r^2-2rx)\,\Bigl[
\frac{1}{16}\bigl\{4 x (3-5 x^2)+r (3-30 x^2+35 x^4)\bigr\}\hatk_i\hatk_j 
\nonumber
\\
&+\frac{1}{48} \bigl(r-4 x+6 r x^2+12 x^3-15 r x^4\bigr) \,\deltaK_{i,j}
\Bigr],
\label{eq:tilde_alpha_ij4}
\end{align}
for $\tilde{\alpha}_{ij}^{(n)}$; and
\begin{align}
  \tilde{\beta}_{ij}^{(2)} & = r^2(-1+x^2)^2\Bigl\{\frac{1}{32}\,  (1-7 x^2) \,\hatk_i\hatk_j + \frac{1}{8}\,(-1+x^2)\,\deltaK_{i,3}\deltaK_{j,3}+ \frac{1}{32}\,(1+x^2)\deltaK_{i,j}\Bigr\}
\label{eq:tilde_beta_ij2}
\\
  \tilde{\beta}_{ij}^{(3)} & =-\frac{r}{16}\,(-1+x^2)^2 \bigl\{-9 x+2 r (-1+7 x^2)\bigr\}\,\Bigl(\hatk_i\deltaK_{j,3}+\hatk_j\deltaK_{i,3}\Bigr),
\label{eq:tilde_beta_ij3}
\\
  \tilde{\beta}_{ij}^{(4)} & =(-1+x^2)\Bigl[
\frac{3}{16}\bigl\{-1+5 x^2+r (9 x-21 x^3)+r^2 (1-14 x^2+21 x^4)\bigr\}\,\hatk_i\hatk_j 
\nonumber
\\
& -\frac{1}{8}(-1+x^2) \bigl\{3-9 r x+r^2 (-1+7 x^2)\bigr\}\,\deltaK_{i,3}\deltaK_{j,3}
\nonumber
\\
& + \frac{1}{16}\,\bigl\{1+3 x^2-3 r (x+3 x^3)+r^2 (-1+2 x^2+7 x^4)\bigr\}\,\deltaK_{i,3}
\Bigr],
\label{eq:tilde_beta_ij4}
\\
  \tilde{\beta}_{ij}^{(5)} & =\frac{1}{16\,r}(-1+x^2)\bigl\{-4 x+6 r (-1+5 x^2)-9 r^2 x (-3+7 x^2)+r^3 (2-28 x^2+42 x^4)\bigr\}\,
\Bigl(\hatk_i\deltaK_{j,3}+\hatk_j\deltaK_{i,3}\Bigr),
\label{eq:tilde_beta_ij5}
\\
  \tilde{\beta}_{ij}^{(6)} & = -\frac{1}{32\,r} \bigl\{8 x (3-5 x^2)+6 r (3-30 x^2+35 x^4)-6 r^2 x (15-70 x^2+63 x^4)
\nonumber
\\
&
+r^3 (-5+105 x^2-315 x^4+231 x^6) \bigr\}\,\hatk_i\hatk_j 
\nonumber
\\
&+\frac{1}{96\,r}\, \bigl\{8 x-24 x^3+6 r^2 x (3+10 x^2-21 x^4)+6 r (-1-6 x^2+15 x^4)
\nonumber
\\
&
+r^3 (3-15 x^2-35 x^4+63 x^6)\bigr\} \,\deltaK_{i,j}
\label{eq:tilde_beta_ij6}
\end{align}
for $\tilde{\beta}_{ij}^{(n)}$.
 
\end{widetext}

Similarly to the $A_{ij}$ term, the traceless conditions also hold for the $B_{ij}$ term, i.e., $\mbox{Tr}[B_{ij}]=0$. To check if it is the case, the following relations are useful:
\begin{align}
&{\rm Tr}[\alpha_{ij}^{(3)}\mu^3 +  \alpha_{ij}^{(4)}\mu^4]=0,
\nonumber
\\
&{\rm Tr}[\beta_{ij}^{(3)}\mu^3 +  \beta_{ij}^{(4)}\mu^4]=0, \quad
{\rm Tr}[\beta_{ij}^{(5)}\mu^5 +  \beta_{ij}^{(6)}\mu^6]=0,
\nonumber
\\
&{\rm Tr}[\tilde{\alpha}_{ij}^{(3)}\mu^3 +  \tilde{\alpha}_{ij}^{(4)}\mu^4]=0,
\nonumber
\\
&{\rm Tr}[\tilde{\beta}_{ij}^{(3)}\mu^3 +  \tilde{\beta}_{ij}^{(4)}\mu^4]=0,\quad
{\rm Tr}[\tilde{\beta}_{ij}^{(5)}\mu^5 +  \tilde{\beta}_{ij}^{(6)}\mu^6]=0.
\end{align}

\section{Analytical expressions of $A_{\rm E}$ and $B_{\rm E}$ terms}
\label{appendix:formulas_AEterm_BEterm}

In this Appendix, the analytical form of the $A_{\rm E}$ and $B_{\rm E}$ terms, given in the expression of the E-mode GI cross power spectrum, i.e., $P_{\rm gE}^{\rm(S)}$ [see Eq.~\eqref{eq:TKO_GI_gE}], is derived based on the formulas summarized in Appendix \ref{appendix:formulas_Aterm_Bterm}.  

The $A_{\rm E}$ and $B_{\rm E}$ terms are related to the $A_{ij}$ and $B_{ij}$ terms through the E/B-mode decomposition at Eq.~\eqref{eq:pk_E/B-mode_decomp}. To be more explicit, they are obtained by substituting the expressions of $A_{ij}$ and $B_{ij}$ into the left-hand side of the following equation:
\begin{align}
    &\bigl(\mathcal{R}_{xx}-\mathcal{R}_{yy}\bigr)\cos(2\phi)+2\mathcal{R}_{xy}\,\sin(2\phi) 
    =  \,\bK\,(1-\mu^2) \mathcal{R}_{\rm E}
    \label{eq:E/B_decomposition_A_B_terms}
\end{align}
with $\mathcal{R}$ being $A$ or $B$. Below, we separately present the resultant expressions for $A_{\rm E}$ and $B_{\rm E}$ in Appendixes \ref{subsec:formulas_AE} and \ref{subsec:formulas_BE}, respectively.

\subsection{$A_{\rm E}$ term}
\label{subsec:formulas_AE}

After applying the E-/B-mode decomposition in Eq.~\eqref{eq:E/B_decomposition_A_B_terms}, the IR-safe integral form of the $A_{\rm E}$ term is given below:
\begin{align} A_{\rm E}(k,\mu)&=\sum_{n} \sum_{\rm X=g,\theta}\mu^n\,\frac{k^3}{4\pi^2}\int_0^\infty dr \int_{-1}^1 dx\,
\nonumber
\\
& \times
\Bigl[ a_{\rm X,E}^{(n)} B_{\theta X\delta} (\bfp,\bfk-\bfp,-\bfk)
\nonumber
\\
&+ \tilde{a}_{\rm X,E}^{(n)} B_{\theta X\delta} (\bfp,-\bfk,\bfk-\bfp)
\nonumber
\\
&\qquad+  b_{\rm X,E}^{(n)} B_{\theta {\rm X}\delta} (\bfk-\bfp,\bfp,-\bfk)
\nonumber
\\
&+  \tilde{b}_{\rm X,E}^{(n)} B_{\theta {\rm X}\delta} (\bfk-\bfp,-\bfk,\bfp)
\Bigr],
\label{eq:AE_term_IR-safe}
\end{align}
with the nonvanishing coefficients, $a_{\rm X,E}^{(n)}$, $\tilde{a}_{\rm X,E}^{(n)}$, $b_{\rm X,E}^{(n)}$, and $\tilde{b}_{\rm X,E}^{(n)}$, summarized as follows:
\begin{align}
 a_{\rm g,E}^{(2)} &= f\,\frac{x}{2r},
\\
 a_{\theta,{\rm E}}^{(2)} &= -f^2\,\frac{(-2+3 r x) (-1+x^2)}{4(1+r^2-2 r x)},
\\
 a_{\theta,{\rm E}}^{(4)} &= f\,\tilde{a}_{\rm g, E}^{(2)},
\\
 \tilde{a}_{\rm g,E}^{(2)} &= f\,\frac{2 x+r \{2-6 x^2+r x (-3+5 x^2)\}}{4 r (1+r^2-2 r x)},
\\
 \tilde{a}_{\theta,{\rm E}}^{(4)} &= f\,\tilde{a}_{\rm g, E}^{(2)},
\end{align}
for $a_{\rm X,E}^{(n)}$ and $\tilde{a}_{\rm X,E}^{(n)}$, and 
\begin{align}
 b_{\rm g,E}^{(2)}&= f\,\frac{1-rx}{2(1+r^2-2rx)},
\\
 b_{\theta,{\rm E}}^{(2)} &= f^2\frac{(-1+3 r x) (-1+x^2)}{4 (1+r^2-2 r x)},
\\
 b_{\theta,{\rm E}}^{(4)} &= f\,\tilde{b}_{\rm g,E}^{(2)},
\\
 \tilde{b}_{\rm g,E}^{(2)}&=f\,\frac{-1+3 x^2+rx(3-5x^2)}{4 (1+r^2-2 r x)},
\\
 \tilde{b}_{\theta,{\rm E}}^{(4)}&=f\,\tilde{b}_{\rm g,E}^{(2)}.
\end{align}
for $b_{\rm X,E}^{(n)}$ and $\tilde{b}_{\rm X,E}^{(n)}$.

\subsection{$B_{\rm E}$ term}
\label{subsec:formulas_BE}

Likewise, one can decompose the $B_{ij}$ term into the E- and B-mode contributions. While the resultant B-mode contribution is shown to be zero, the E-mode contributions become nonvanishing, and are expressed as follows:
\begin{align}
 B_{\rm E}(k,\mu)&=\sum_{n} \mu^n\,\frac{k^3}{4\pi^2}\,\int_0^\infty dr\int_{-1}^1 dx\,
\nonumber
\\
&\quad\times
\Biggl[\,\Bigl\{f^2\, \alpha_{\rm E}^{(n)} P_{{\rm g}\theta} (k r) 
+ f^3\,\beta_{\rm E}^{(n)} P_{\theta\theta} (k r) 
\Bigr\}\,
\nonumber
\\
&\times\frac{P_{\theta\delta}(k\sqrt{1+r^2-2rx})}{(1+r^2-2rx)^2}
\nonumber
\\
&
+\Bigl\{f^2\, \tilde{\alpha}_{\rm E}^{(n)} P_{{\rm g}\theta} (k\sqrt{1+r^2-2rx} ) 
\nonumber
\\
&+ f^3\,\tilde{\beta}_{\rm E}^{(n)} P_{\theta\theta} (kk\sqrt{1+r^2-2rx}) 
\Bigr\}\,
\nonumber
\\
&\times\frac{P_{\theta\delta}(k\,r)}{(1+r^2-2rx)^2}
\Biggr].
\label{eq:BE_term}
\end{align}
The nonvanishing coefficients in the integral, $\alpha_{\rm E}^{(n)}$, $\tilde{\alpha}_{\rm E}^{(n)}$, $\beta_{\rm E}^{(n)}$, and $\tilde{\beta}_{\rm E}^{(n)}$, are summarized below:
\begin{align}
 \alpha_{\rm E}^{(2)} &=\frac{1}{16}(-1+x^2) \{4-8 r x+r^2 (-1+5 x^2)\},
\\
 \alpha_{\rm E}^{(4)} &=\frac{1}{16r}\bigl[8 x+r \{12-36 x^2+12r x (-3+5 x^2)
\nonumber
\\
& +r^2 (-3+30 x^2-35 x^4)\}\bigr],
\end{align}
for $\alpha_{\rm E}^{(n)}$;
\begin{align}
 \beta_{\rm E}^{(2)} &=-\frac{1}{32}(-1+x^2)^2 \{6-12 r x+r^2 (-1+7 x^2)\},
\\
 \beta_{\rm E}^{(4)} &=\frac{3}{16r}(-1+x^2)\bigl\{-4 x+4 r (-1+5 x^2)
\nonumber
\\
&-5 r^2 x (-3+7 x^2)+r^3 (1-14 x^2+21 x^4)\bigr\},
\\
 \beta_{\rm E}^{(6)} &=\frac{1}{32r}\,\bigl[
8 x (-3+5 x^2)-6 r  (3-30 x^2+35 x^4)
\nonumber
\\
&+6 r^2x (15-70 x^2+63 x^4)
\nonumber
\\
&+r^3 (5-105 x^2+315 x^4-231 x^6)\bigr],
\end{align}
for $\beta_{\rm E}^{(n)}$;
\begin{align}
 \tilde{\alpha}_{\rm E}^{(2)} &=\frac{1}{16}(1+r^2-2 r x) (1-6 x^2+5 x^4),
\\
 \tilde{\alpha}_{\rm E}^{(4)} &=-\frac{1}{16r}\,(1+r^2-2 r x) \{4x(3-5x^2)
\nonumber
\\
&+r(3-30x^2+35x^4)\},
\end{align}
for $\tilde{\alpha}_{\rm E}^{(n)}$; and  
\begin{align}
 \tilde{\beta}_{\rm E}^{(2)} &= -\frac{1}{32}\,r^2(-1+x^2)^2 (-1+7 x^2),
\\
 \tilde{\beta}_{\rm E}^{(4)} &= \frac{3}{16}\,(-1+x^2)\bigl\{-1+5 x^2+ r (9x-21 x^3)
\nonumber
\\
&+r^2 (1-14 x^2+21 x^4)\bigr\},
\\
  \tilde{\beta}_{\rm E}^{(6)} &= \frac{1}{32r}\,\bigl\{
8 x (-3+5 x^2)-6 r (3-30 x^2+35 x^4)
\nonumber
\\
&+6 r^2 x (15-70 x^2+63 x^4)
\nonumber
\\
 &+r^3 (5-105 x^2+315 x^4-231 x^6)\bigr\},
\end{align}
for $\tilde{\beta}_{\rm E}^{(n)}$.

\section{On the Gaussian linear covariance}
\label{appendix:validity_covariance}

In Sec.~\ref{subsec:inference}, the parameter estimation study using the GI cross power spectrum was examined adopting the analytical Gaussian covariance computed with linear theory. In general, for the scales where the linear theory breaks down, the non-Gaussianity of the covariance comes to play a role, and this can not only lead to the nonzero off diagonal components between different Fourier modes but also change the diagonal components (e.g., see ~\cite{Takahashi_etal2009,Blot2015} for numerical studies and \cite{Digvijay_Scoccimarro2019,Sugiyama_etal2020,Taruya_Nishimichi_Jeong2021} for perturbation theory approaches). 

\begin{figure*}[tb]
 \includegraphics[width=12cm,angle=0]{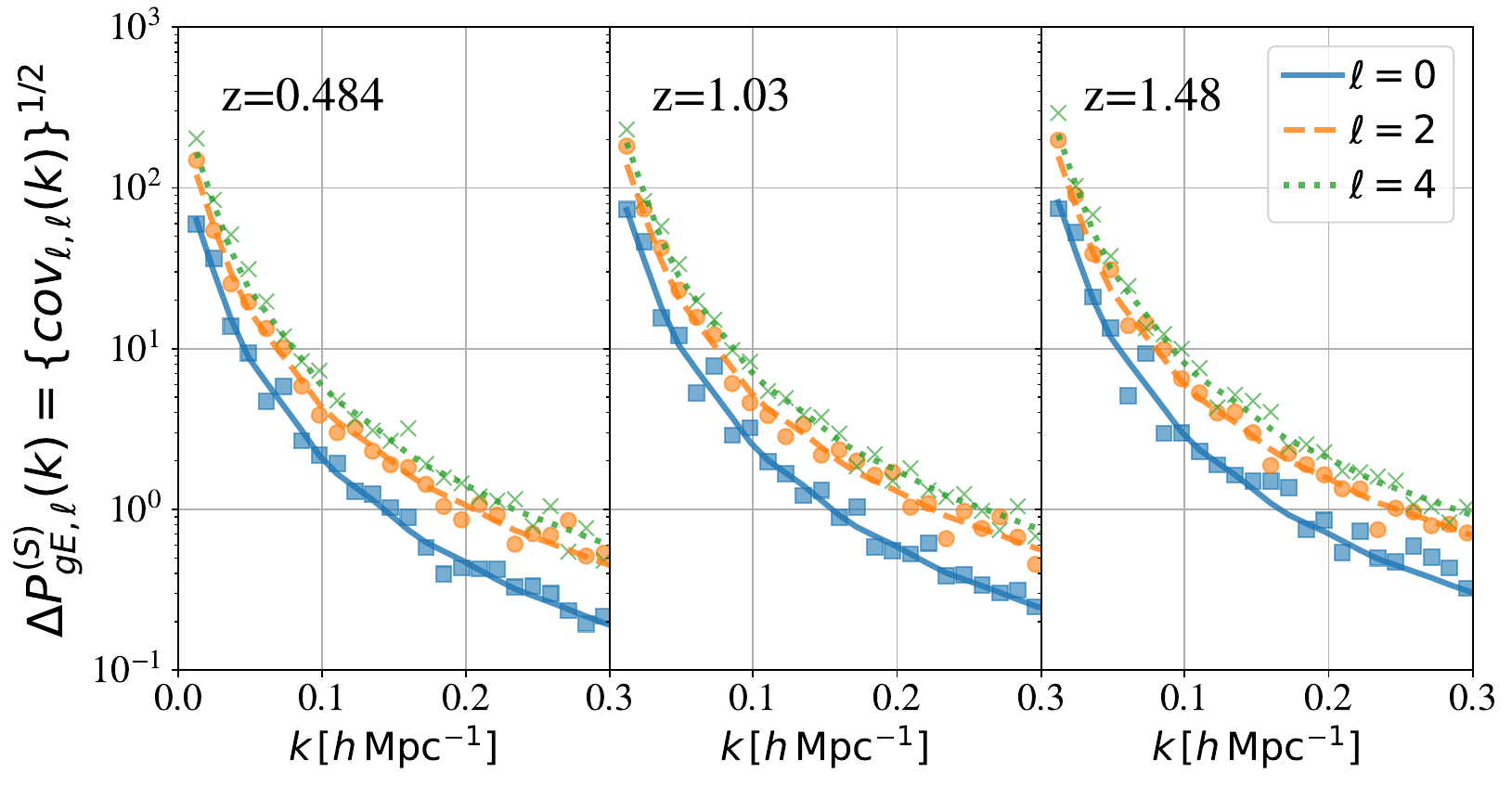}
\caption{Square of the covariance matrix for the power spectrum multipoles $P_{\rm gE,\ell}^{\rm(S)}$ at $z=0.484$ (left), $1.03$ (middle), and $1.48$ (right). The error of the power spectrum mean measured from simulations, $\Delta P_{\rm gE,\ell}^{\rm(S)}$, is compared to the predictions of linear theory formula for the Gaussian covariance [see Eq.~\eqref{eq:Gaussian_cov} with \eqref{eq:linear_pk}]. In computing the linear theory predictions, the scale-dependent nature of the linear bias parameters, $b_1$ and $b_{\rm K}$, as shown in Fig.~\ref{fig:scale-dept_bias}, is incorporated into the analytical formulas.  
\label{fig:cov}
}
\end{figure*}

Using the $20$ realizations of the $N$-body simulations, we checked the validity of the Gaussian linear covariance by directly comparing it with the measured covariance. 
Figure~\ref{fig:cov} plots the square root of the diagonal component of the covariance, ${\rm cov}_{\ell,\ell}^{1/2}$, as a function of $k$, corresponding to the standard error of the mean power spectra over the total volume of $V_{\rm s}=20\,h^{-3}$\,Gpc$^3$. While the symbols are measured from simulations, continuous curves are the results from the Gaussian linear covariance at Eq.~\eqref{eq:Gaussian_cov}, in which the number of Fourier modes in each $k$ bin $N_{\rm k}$, number density of halos $n_{\rm g}$, and the scatter of the intrinsic ellipticity $\sigma_{\gamma}$ are those measured from the simulation data (see Table~\ref{tab:halo_num_shape}).

Figure~\ref{fig:cov} shows that the Gaussian linear covariance provides a very good approximation to the covariance of $P^{\rm(S)}_{\rm gE,\ell}$ even at $k\sim0.3\,h$\,Mpc$^{-1}$, 
where the one-loop PT prediction usually becomes inadequate. 
Owing to the limited number of realizations, we could not check with simulations the off-diagonal components between different Fourier modes or multipoles. Nevertheless, in general, non-Gaussian contributions impact the off diagonal covariance at the scale where the diagonal part deviates from the Gaussian linear covariance \cite{Taruya_Nishimichi_Jeong2021}.
 In this respect, as long as we stick to the scales of $k\lesssim0.3\,h$\,Mpc$^{-1}$, the Gaussian linear covariance still gives a reasonable approximation to the actual power spectrum covariance, and one can apply it to the parameter estimation study. 

\end{document}